\documentclass[sigconf, nonacm]{acmart}

\usepackage[ruled,linesnumbered,noend]{algorithm2e}
\usepackage{color}
\usepackage{setspace}
\usepackage{multirow}
\usepackage{makecell}
\usepackage{enumitem}
\usepackage{amsmath}

\AtBeginDocument{%
  }

\begin{document}

\title[cuRPQ]{cuRPQ: A High-Performance GPU-Based Framework for Processing Regular and Conjunctive Regular Path Queries}

\author{Sungwoo Park}
\orcid{0009-0008-3927-7935}
\affiliation{%
  \institution{Korea Advanced Institute of Science and Technology}
  \city{Daejeon}
  \country{Republic of Korea}
}
\email{sungwoo.park@kaist.ac.kr}

\author{Seohyeon Kim}
\orcid{0009-0005-7944-8469}
\affiliation{%
  \institution{Korea Advanced Institute of Science and Technology}
  \city{Daejeon}
  \country{Republic of Korea}
}
\email{asdf0322@kaist.ac.kr}

\author{Min-Soo Kim}
\authornote{Corresponding author.}
\orcid{0000-0002-7852-0853}
\affiliation{%
  \institution{Korea Advanced Institute of Science and Technology}
  \city{Daejeon}
  \country{Republic of Korea}
}
\email{minsoo.k@kaist.ac.kr}


\begin{abstract}

Regular path queries\,(RPQs) are fundamental for path-constrained reachability analysis, and more complex variants such as conjunctive regular path queries\,(CRPQs) are increasingly used in graph analytics.
Evaluating these queries is computationally expensive, but to the best of our knowledge, no prior work has explored GPU acceleration.
In this paper, we propose cuRPQ, a high-performance GPU-optimized framework for processing RPQs and CRPQs.
cuRPQ addresses the key GPU challenges through a novel traversal algorithm, an efficient visited-set management scheme, and a concurrent exploration-materialization strategy.
Extensive experiments show that cuRPQ outperforms state-of-the-art methods by orders of magnitude, without out-of-memory errors.

\end{abstract}





\maketitle

\noindent\textbf{Author's version.}
This paper has been accepted for publication in the Proceedings of the ACM SIGMOD International Conference on Management of Data\,(SIGMOD 2026).
\vspace{0.5em}

\section{Introduction}
\label{sec:introduction}

\begin{figure}[t]
    \vspace*{-0.05cm}
    \centerline{\includegraphics[width=3.45in]{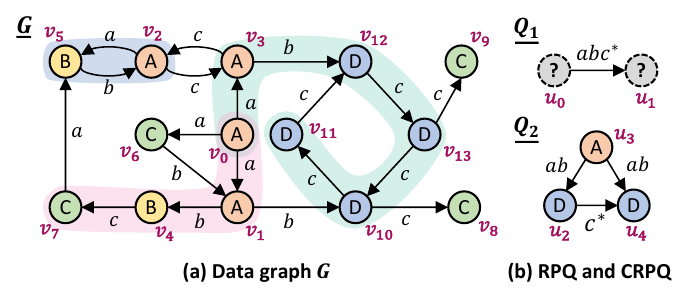}}
    \vspace*{-0.4cm}
    \caption{Example of RPQ and CRPQ on a data graph.}
    \label{fig:RPQ}
    \vspace*{-0.55cm}
\end{figure}

A \textit{Regular Path Query}\,(RPQ) is one of the most expressive and powerful graph queries~\cite{wadhwa2019efficiently, faltin2023distributed}, which finds all pairs of vertices connected by a path whose sequence of edge labels matches a given regular expression.
RPQs have two key characteristics: they allow paths of \textbf{arbitrary length} and return \textbf{distinct vertex pairs}~\cite{angles2017foundations, martens2019dichotomies}.
For example, given a data graph $G$ with vertex and edge labels as in Figure~\ref{fig:RPQ}, an RPQ $Q_1$ is the task of finding vertex pairs that satisfy $u_0$ $\xrightarrow{abc^*}$ $u_1$.
It returns a total of 13 results\footnote{$\{(v_0, v_1),$ $(v_0, v_4),$ $(v_0, v_7),$ $(v_0, v_8),$ $(v_0, v_9),$ $(v_0, v_{10}),$ $(v_0, v_{11}),$ $(v_0, v_{12}),$ $(v_0, v_{13}),$ $(v_2, v_2),$ $(v_2, v_3),$ $(v_7, v_2),$ $(v_7, v_3)\}$}, among which $\{(v_2, v_2),$ $(v_0, v_7),$ $(v_0, v_{11})\}$ correspond to the paths $ab$, $abc$, and $abccc$, respectively.
$Q_1$ represents an \textit{all-pairs RPQ}, where both endpoints are variables, whereas the RPQ can also be expressed as a \textit{single-source RPQ} when one of the endpoints is fixed to a constant vertex~\cite{belyanin2024single, davoust2016processing}.
RPQs are applied in various domains that require reachability analysis with path constraints, such as social network analysis~\cite{konstas2009social}, fraud detection~\cite{xu2022efficiently}, network security~\cite{phillips1998graph}, and bioinformatics~\cite{leser2005query, van2000representing}.
Given their widespread applications, many graph query languages --- such as openCypher~\cite{green2018opencypher}, SPARQL~\cite{SPARQL2013}, and ISO/GQL~\cite{deutsch2022graph} --- support RPQs~\cite{gheerbrant2025database, pang2024materialized} as a fundamental operation.

Recently, as analyses required in various applications have become more complex~\cite{szarnyas2022ldbc, poulovassilis2016approximation, xu2022efficiently, dorpinghaus2022context, dorpinghaus2022social}, the demand for \textit{Conjunctive Regular Path Queries}\,(CRPQ)~\cite{cruz1987graphical} has increased.
A CRPQ finds subgraph patterns that satisfy the regular expression constraints of multiple RPQ atoms simultaneously.
In Figure~\ref{fig:RPQ}, $Q_2$ on $(u_2, u_3, u_4)$ maps to $\{(v_{10}, v_0, v_{10}), (v_{10}, v_0, v_{12}), (v_{12}, v_0, v_{10}), (v_{12}, v_0, v_{12})\}$.
Representatively, CRPQs are applied in information propagation analysis to trace the creator User\,($u$) and the related Post\,($p$) of a Message\,($m$) with a specific Tag\,($t$:$Sports$)~\cite{szarnyas2022ldbc}.
This can be expressed as a CRPQ that simultaneously satisfies three RPQ atoms: $m \xrightarrow{hasTag} t$:$Sports$, $m \xrightarrow{hasCreator} u$, and $m \xrightarrow{replyOf^*} p$.
The second and third atoms correspond to all-pairs RPQs, since both endpoints are variables, and such all-pairs RPQs are frequently evaluated in CRPQs~\cite{szarnyas2022ldbc}. 
In general, a CRPQ is processed by evaluating each RPQ atom and then combining the results using joins or a \textit{worst-case optimal multi-way join}\,(WCOJ)~\cite{veldhuizen2014leapfrog, ngo2012worst, karalis2024efficient, cucumides2023size, figueira2020containment}.
However, when evaluating all-pairs RPQs with densely connected edge labels, such as the $replyOf$ relation among messages~\cite{ugander2011anatomy}, the computational cost can become prohibitively high, since paths between almost all vertex pairs must be considered.

Most existing studies have addressed RPQs using two main approaches: (1) the algebra-based approach and (2) the automata-based approach~\cite{pang2024materialized, arroyuelo2024optimizing, yakovets2016query, arroyuelo2025evaluating, gou2024lm, bonifati2022querying, miura2019accelerating}. 
The algebra-based approach~\cite{neumann2020umbra, raasveldt2019duckdb} computes the transitive closure\,(e.g., $c^*$) by repeatedly applying the $\alpha$-operator~\cite{agrawal2002alpha}, which iteratively performs join operations until a fixed point is reached. 
Other regular expression operators, such as concatenation\,(e.g., $ab$) and alternation\,(e.g., $a+b$), are handled through relational algebra operations such as join, union, and distinct. 
In contrast, the automata-based approach~\cite{gou2024lm, pacaci2020regular, arroyuelo2024optimizing, koschmieder2012regular, nguyen2017efficient} simultaneously performs data-graph traversal and automaton transitions corresponding to the given regular expression.
For each starting vertex, a \textit{visited set} is maintained, storing pairs of data graph vertices and automaton states to ensure that only non-duplicate paths are explored.

In general, the computational complexity of all-pairs RPQ evaluation is \textit{cubic} in the number of vertices for algebra-based methods, \textit{quadratic} for automata-based methods, and \textit{NP-complete} for CRPQs~\cite{
barcelo2013querying, agrawal1987direct, abo2025output, barcelo2017graph, casel2023fine, barcelo2012expressive}.
These complexities represent a major bottleneck, motivating the use of GPUs to exploit their massive parallelism.
Although GPU-based acceleration has been extensively studied for graph query processing~\cite{park2025cumatch, lai2022accelerating, wei2022stmatch, guo2020gpu, sun2023efficient, zeng2020gsi, xiang2021cuts, yuan2024faster}, to the best of our knowledge, no prior work has proposed a GPU-based method for RPQ processing.
Designing such a GPU-based method for RPQ and CRPQ evaluation introduces \textbf{three fundamental challenges}.

\textbf{First,} naive GPU-based methods for arbitrary-length RPQs can 
suffer from severe \textit{performance degradation} and may produce \textit{incorrect answers} due to \textit{traversal path length limitations}.
Because of limited GPU memory, most GPU-based graph processing methods adopt DFS-based or DFS-like traversal strategies~\cite{park2025cumatch, oh2025gflux, bisson2017high, wei2022stmatch, yuan2024faster, xiang2021cuts, lin2016network, lai2022accelerating, sun2023efficient}. 
However, DFS inherently explores long paths first, often checking reachability inefficiently by traversing unnecessarily deep paths.
On GPUs, where only a portion of the graph can be loaded at a time, traversal depth is strictly limited, and DFS may easily exceed this limit, failing to discover valid paths.

\textbf{Second,} naive GPU-based methods significantly \textit{reduce parallelism as the visited set quickly exceeds} GPU memory capacity.
Achieving high performance on GPUs requires simultaneous traversal from thousands of starting vertices, but the size of the visited set grows proportionally with the number of starting vertices and can easily surpass available GPU memory~\cite{arroyuelo2024optimizing}. 
This constraint prevents large-scale parallel traversal, reducing GPU utilization and thereby degrading overall performance.

\textbf{Third,} naive GPU-based methods \textit{cannot evaluate CRPQs, as the materialization of RPQ results often exceeds} GPU memory capacity, and managing result explosion is especially challenging for all-pairs RPQs.
In addition, the execution time of an all-pairs RPQ can vary greatly depending on optimization strategies\,(e.g., join ordering and materialized views in algebra-based approaches).
To enable such optimizations in automata-based approaches, it is necessary to employ state-of-the-art plan representations such as WavePlan~\cite{yakovets2016query}, which extend automaton-based plans, but still require the materialization of sub-RPQs, whose results can grow explosively.
This problem also occurs in GPU-based DBMSs that support materialization~\cite{park2025cumatch}; for instance, the state-of-the-art GPU-based DBMS HeavyDB~\cite{heavydb2025} and GPU library RAPIDS~\cite{RAPIDS2025} are only feasible when the materialized RPQ results fit entirely within GPU memory.

To address the above challenges, we propose \textbf{cuRPQ, a high-performance GPU-based framework} for processing RPQs and CRPQs.
It efficiently handles computationally expensive queries, such as all-pairs RPQs through a series of GPU-oriented optimizations, and can be seamlessly extended to support CRPQs.
For the first challenge, we propose a \textbf{hop-limited level-wise DFS} method that performs DFS within fixed-hop ranges and advances the search in a level-wise manner.
This design preserves the low memory footprint of DFS while achieving BFS-like progression across levels.
For the second challenge, we propose an \textbf{on-demand segment pooling} technique that manages the visited set using fine-grained segments. This enables dynamic memory allocation and release, sustaining large-scale parallel exploration without exhausting GPU memory capacity.
For the third challenge, we propose a \textbf{concurrent exploration-materialization} strategy for heterogeneous GPU-CPU environments, where GPUs focus on path exploration while CPUs concurrently perform GPU-aware graph materialization of large RPQ results.
Through these techniques, cuRPQ effectively addresses the key challenges of GPU-based RPQ and CRPQ processing, fully exploiting GPU parallelism to achieve high performance and scalability.

The main contributions of this paper are as follows:
\begin{itemize}[labelindent=0em,labelsep=0.3em,leftmargin=*,itemsep=0em]
    \item We propose cuRPQ, a GPU-optimized framework for processing RPQ/CRPQ queries.
    \item We propose the hop-limited level-wise DFS, a novel traversal method for efficiently exploring arbitrary-length paths while maintaining low memory overhead.
    \item We propose the on-demand segment pooling technique, a fine-grained memory management technique that alleviates GPU memory bottlenecks in managing the visited set.
    \item We propose the concurrent exploration-materialization strategy that fully exploits GPU parallelism beyond GPU memory limits.
    \item Extensive\,experiments\,show\,that\,cuRPQ outperforms state-of-the-art CPU-based algebra and automata methods by\,up\,to\,4,945X and 269X, respectively, even for queries producing trillions of results.
\end{itemize}

\vspace{-0.1cm}
\section{Preliminaries}
\label{sec:preliminaries}

\subsection{Problem Definition}
\label{sec:problem_definition}

We consider a directed graph $G = (V, E, L)$ in which both vertices and edges are labeled. Here, $V$ denotes the set of vertices, $E$ the set of edges, and $L$ the labeling function that assigns labels to vertices and edges.
We denote the label of a vertex $v_i$ by $L(v_i)$ and the label of an edge $(v_i, v_j)$ by $L(v_i, v_j)$. 
We define RPQ in Definition~\ref{definition:rpq}~\cite{gou2024lm, pacaci2020regular, arroyuelo2025evaluating, arroyuelo2024optimizing}.

\begin{definition}
\label{definition:rpq}
\textbf{Regular Path Query\,(RPQ)}:
Given a data graph $G=(V,E,L)$, an RPQ is denoted as $x \xrightarrow{\rho} y$, where $x$ and $y$ are query variables that can take values from $V$, and $\rho$ is a regular expression over the set of edge labels of $G$. 
The result of the RPQ is the set of all pairs $(x, y) \in V \times V$ such that there exists a path $p = (v_i \rightarrow v_{i+1} \rightarrow \cdots \rightarrow v_j)$ in $G$ whose edge-label sequence $L(v_i, v_{i+1})\, L(v_{i+1}, v_{i+2})\, \cdots\, L(v_{j-1}, v_j)$ matches the regular expression $\rho$, where $x$ = $v_i$ and $y$ = $v_j$.
\end{definition}

By definition, an RPQ returns distinct (start, end) vertex pairs connected by at least one path matching a given regular expression.
This pair-based semantics is standard in the RPQ literature and sufficient for many practical use cases that primarily focus on reachability~\cite{mendelzon1995finding}.
A CRPQ generalizes RPQs by allowing multiple RPQ atoms to be combined in a conjunction. 
We define CRPQ in Definition~\ref{definition:crpq}~\cite{figueira2023conjunctive, figueira2025minimizing, figueira2020containment}.

\begin{definition}
\label{definition:crpq}
\textbf{Conjunctive Regular Path Query (CRPQ)}:
Given a data graph $G=(V, E, L)$ and $Q=(V_q, E_q, L_q)$, where each edge in $E_q$ 
is of the form $x \xrightarrow{\rho} y$ with $x,y \in V_q$ and a regular expression $\rho$. 
The result of the CRPQ is the set of all possible homomorphisms $f: V_q \rightarrow V$ such that (1) $\forall u \in V_q, L_q(u) = L(f(u))$; and (2) $\forall (x \xrightarrow{\rho} y) \in E_q$, there exists a path from $f(x)$ to $f(y)$ in $G$ whose edge-label sequence matches $\rho$. 
\end{definition}

\vspace{-0.34cm}
\subsection{Processing RPQs}
\label{sec:RPQ}

Graph pattern matching, also known as a \textit{Conjunctive Query}\,(CQ), focuses on structural correspondences between a given pattern and a subgraph of a data graph~\cite{bonifati2022querying, angles2017foundations}.
In contrast, an RPQ is a path query that supports transitive closure operators such as the Kleene star\,($*$), enabling the exploration of arbitrary-length paths, and it returns only distinct vertex pairs even when multiple paths exist between them.

\vspace*{0.1cm}
\textbf{Algebra-based approach}~\cite{neumann2020umbra, raasveldt2019duckdb}: 
This approach employs the $\alpha$-operator~\cite{agrawal2002alpha}, commonly expressed through the \textbf{\texttt{WITH RECURSIVE}} clause.
In RPQs, this operator is used to compute transitive closure\,(e.g., $c^*$).
The computation starts from the relation consisting of edges with the given label\,(e.g., $c$ in $c^*$) and repeatedly joins it to expand reachable vertex pairs one hop at a time.
The newly discovered pairs are merged with the accumulated set of reachable pairs through a union operation to preserve distinct results, and the process continues until no further pairs are discovered.
Non-transitive edge patterns are processed by standard join operations.
When several relations correspond to an edge label, they are first unified before processing.
For example, in $Q_1$ of Figure~\ref{fig:RPQ}, suppose there exist relations corresponding to edge labels $a$, $b$, and $c$.
The relation for edge label $a$ consists of tuples $\{(v_0, v_1), (v_0, v_3), (v_0, v_6), (v_2, v_5), (v_7, v_5)\}$.
Here, $c^*$ is computed as intermediate data through the $\alpha$-operator, and the final results can then be derived by applying join and distinct operations with the relations corresponding to $a$ and $b$.

\begin{figure}[b]
    \centerline{\includegraphics[width=3.4in, trim=4 0 0 0, clip]{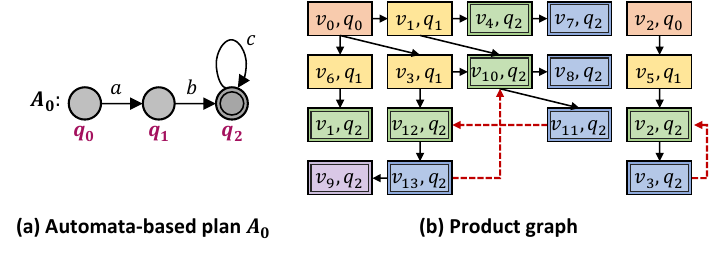}}
    \vspace*{-0.4cm}
    \caption{Example of automata-based plan and product graph.}
    \label{fig:AM_based}
    \vspace*{-0.2cm}
\end{figure}

\vspace*{0.1cm}
\textbf{Automata-based approach}~\cite{gou2024lm, pacaci2020regular, arroyuelo2024optimizing, koschmieder2012regular, nguyen2017efficient}:
This approach exploits the automaton corresponding to the regular expression of the given RPQ.
Traversal starts from the possible starting vertices in the data graph\,(e.g., all vertices in an all-pairs RPQ) and the initial state of the RPQ automaton, simultaneously performing automaton transitions and edge exploration.
Whenever the automaton reaches a final state, the corresponding start-end vertex pair is added to the RPQ result set.
Conceptually, this process can be viewed as operating over the \textit{product graph}~\cite{gou2024lm, pacaci2020regular, arroyuelo2024optimizing}, in which each vertex corresponds to a pair consisting of a data-graph vertex and an automaton state.
For each starting vertex, a \textit{visited set} records explored product graph vertices to avoid redundancy and ensure distinct results.
For example, Figure~\ref{fig:AM_based} presents the automata-based evaluation of $Q_1$ on the data graph $G$ shown in Figure~\ref{fig:RPQ}.
Figure~\ref{fig:AM_based}(a) shows the automata plan corresponding to $abc^*$, while Figure~\ref{fig:AM_based}(b) depicts a portion of the product graph derived from starting at $v_0$ and $v_2$.
The red dashed lines indicate edges where traversal terminates because the corresponding product graph vertex is already in the visited set.
In the traversal process, the vertices marked with double squares corresponding to 2-hop\,(green), 3-hop\,(blue), and 4-hop\,(purple) are included in the RPQ result set.

\vspace{-0.1cm}
\subsection{WavePlan}
\label{sec:waveplan}

\begin{figure}[b]
\vspace*{-0.4cm}
    \centerline{\includegraphics[width=3.3in, trim={0 0 0 5}, clip]{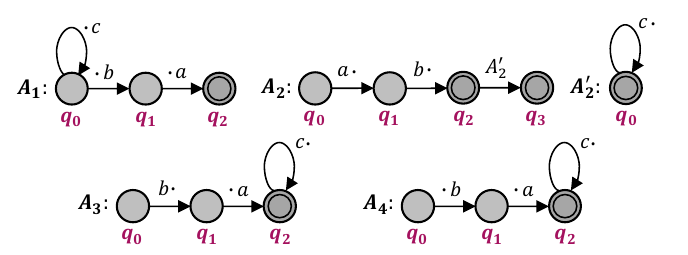}}
    \vspace*{-0.25cm}
    \caption{Example of WavePlans.}
    \label{fig:WavePlan}
    \vspace*{-0.2cm}
\end{figure}

WavePlan~\cite{yakovets2016query} is a plan representation that extends the automata-based plan while incorporating the advantages of the algebra-based approach.
The algebra-based approach materializes the results of transitive closure\,(e.g., $c^*$), which enables the reuse of partial paths and allows for further optimizations such as join ordering~\cite{yakovets2016query}.
In contrast, the automata-based approach follows automaton state transitions and directly passes results to subsequent operations, enabling query processing in a \textit{pipelined fashion}~\cite{yakovets2016query}.
WavePlan primarily represents queries in an automata-based plan, while also allowing sub-RPQ results to be materialized for reuse and supporting diverse exploration orderings as needed.
Figure~\ref{fig:WavePlan} shows several WavePlans for $abc^*$.
Beyond the $A_0$ plan from Figure~\ref{fig:AM_based}, WavePlan supports the \textit{reverse} plan $A_1$, which uses not only out-edges\,(e.g., $a\cdot$) but also in-edges\,(e.g., $\cdot a$); the \textit{loop-cache} plan $A_2$, which reuses results from other automata; and the \textit{start-in-the-middle} plans $A_3$ and $A_4$, which begin traversal from the middle of the path expression.

Finally, WavePlan is a plan representation that integrates the strengths of the automata-based and algebra-based approaches through an automata-based execution plan.
By adopting an automata-based design, cuRPQ naturally integrates with WavePlan and can leverage execution-plan-level optimizations, even in GPU environments with limited memory capacity.

\vspace{-0.1cm}
\subsection{Labeled Grid Format}
\label{sec:LGF}

The \textit{Labeled Grid Format}\,(LGF)~\cite{park2025cumatch} extends the conventional 2D grid format~\cite{ko2018turbograph++, maass2017mosaic, wang2022neutronstar, ma2019neugraph, zhu2015gridgraph, oh2025gflux} --- which partitions source and destination vertex ranges into fixed widths and heights --- from simple graphs to labeled graphs.
In LGF, a separate grid is maintained for each edge label, and the source and destination vertex ranges of each grid are logically partitioned based on vertex labels.
Each partition, called a \textit{block}, contains edges sharing the same source label, destination label, and edge label.
LGF maintains block-level metadata called the \textit{GridMap}.
GridMap follows a 3D array structure, each element in GridMap has coordinates $(x, y, z)$, where $x$ is the \textit{block row ID} for source vertex, $y$ is the \textit{block column ID} for destination vertex, and $z$ is the \textit{grid ID} for edge label.
LGF also maintains a \textit{VertexLabel} table that stores vertex label names and their block-level ranges, and an \textit{EdgeLabel} table that stores edge label names.
Each block can be recursively partitioned into four quadrants along the source and destination dimensions, producing smaller \textit{slices} until their sizes fall below a specific threshold~\cite{park2025cumatch, oh2025gflux}.
This guarantees that all partitions\,(i.e., slices) remain below the threshold, preventing any single slice from growing too large to fit in GPU memory.
Each slice is stored in an adjacency-list representation\,(e.g., Compressed Sparse Row~\cite{gilbert1992sparse}, trie-based index), and distinct slices are maintained for out-edges and in-edges to support bidirectional traversal.
Finally, LGF can operate efficiently in both in-memory and disk-based environments~\cite{park2025cumatch, oh2025gflux}.

\begin{figure}[b]
\vspace*{-0.45cm}
    \centerline{\includegraphics[width=3.4in]{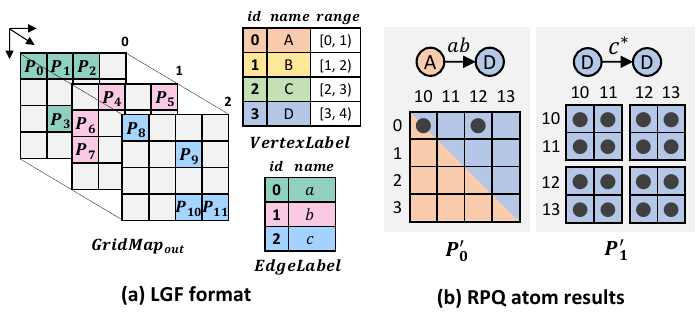}}
    \vspace*{-0.4cm}
    \caption{Example of Labeled Grid Format\,(LGF).}
    \label{fig:LGF}
    \vspace*{-0.1cm}
\end{figure}

Figure~\ref{fig:LGF}(a) shows an example of LGF for the data graph $G$ in Figure~\ref{fig:RPQ}, which consists of four vertex labels $\{A, B, C, D\}$ and three edge labels $\{a, b, c\}$.
For simplicity, all vertices with the same vertex label are assumed to fit within a single block\,($P_i$) (i.e., both block width and height are 4), and each block is assumed to consist of a single slice\,($S_i$).
Thus, all ranges in the VertexLabel table have a length one\,(e.g., [0, 1)).
Including in-edge slices, there are 24 blocks and slices in total, but due to space constraints only $GridMap_{out}$ for out-edge slices is shown.
The edges stored in each slice are listed in Table~\ref{tab:LGF}.
$P_5$, pointed to by $(0, 3, 1)$ in $GridMap$, corresponds to a single slice $S_5$ and contains two edges $\{(v_1, v_{10}), (v_3, v_{12})\}$ whose source vertex label is $A$, destination vertex label is $D$, and edge label is $b$.

In this paper, we adopt the LGF format for processing RPQs and CRPQs.
In the automata-based approach, efficient traversal by edge label is required.
LGF supports this by maintaining a separate grid for each label, enabling direct access to the edge set corresponding to a given edge label\,(i.e., the slices within a single grid).
Moreover, since LGF stores both out-edge and in-edge slices, it supports bidirectional traversal and is well suited for the reverse plans required by WavePlan.
In particular, subdividing blocks into smaller slices enables LGF to materialize RPQ results in a partitioned form, ensuring that even the potentially huge result sizes of all-pairs RPQs can be stored slice by slice.
For example, Figure~\ref{fig:LGF}(b) shows the results of two RPQ atoms required for $Q_2$ in Figure~\ref{fig:RPQ}.
$P'_0$ and $P'_1$ correspond to the RPQ results of $ab$ and $c^*$, respectively, where $P'_0$ contains two edges $\{(v_0, v_{10}), (v_0, v_{12})\}$.
If the threshold is limited to four edges per slice, block $P'_1$ is subdivided into four quadrants along the source and destination dimensions, resulting in four slices.

\begin{table}[b]
\vspace*{-0.25cm}
\setlength{\tabcolsep}{2.2pt}
\caption{Example of edges stored in each slice.}
\vspace*{-0.25cm}
\label{tab:LGF}
\small
\begin{tabular}{c|l}
\toprule
\textbf{Edge labels} & \multicolumn{1}{c}{\textbf{Edges per slice}} \\
\midrule
$\boldsymbol{a}$ &
$\boldsymbol{S_0}$: $(v_0, v_1), (v_0, v_3)$,
$\boldsymbol{S_1}$: $(v_2, v_5)$,
$\boldsymbol{S_2}$: $(v_0, v_6)$,
$\boldsymbol{S_3}$: $(v_7, v_5)$ \\
\midrule
$\boldsymbol{b}$ &
$\boldsymbol{S_4}$: $(v_1, v_4)$,
$\boldsymbol{S_5}$: $(v_1, v_{10}), (v_3, v_{12})$,
$\boldsymbol{S_6}$: $(v_5, v_2)$,
$\boldsymbol{S_7}$: $(v_6, v_1)$ \\
\midrule
\multirow{2}{*}{$\boldsymbol{c}$} &
$\boldsymbol{S_8}$: $(v_2, v_3), (v_3, v_2)$,
$\boldsymbol{S_9}$: $(v_4, v_7)$,
$\boldsymbol{S_{10}}$: $(v_{10}, v_8), (v_{13}, v_9)$ \\
& $\boldsymbol{S_{11}}$: $(v_{10}, v_{11}), (v_{11}, v_{12}), (v_{12}, v_{13}), (v_{13}, v_{10})$ \\
\bottomrule
\end{tabular}
\vspace*{-0.05cm}
\end{table}

\vspace{-0.15cm}
\section{Challenges of GPU-Based CRPQ Processing}
\label{sec:challenges}

In this section, we elaborate on the challenges highlighted in the Introduction, based on the discussion in Section~\ref{sec:preliminaries}.

\vspace*{0.1cm}
\textbf{Challenge 1\,(Incorrect answers caused by traversal path length limitations):}
Most GPU-based graph query processing methods~\cite{park2025cumatch, oh2025gflux, bisson2017high, wei2022stmatch, yuan2024faster, xiang2021cuts, lin2016network, lai2022accelerating, sun2023efficient} adopt DFS-based traversal strategies due to the limited capacity of GPU and shared memory. 
Some hybrid methods~\cite{park2025cumatch, xiang2021cuts, lin2016network, lai2022accelerating, sun2023efficient} partially incorporate BFS, but GPU memory constraints for storing all frontier vertices prevent full BFS traversal~\cite{xiang2021cuts, sun2023efficient}, leaving them essentially DFS-based.
In detail, as defined in Definition~\ref{definition:rpq}, evaluating an RPQ requires exploring paths starting from $x=v_i$ that satisfy the regular expression $\rho$.
On GPUs, this exploration is typically performed in a depth-first manner along with automaton transitions.
However, DFS tends to explore longer paths first, which leads to inefficiency by unnecessarily expanding paths even when shorter valid ones exist. 
For example, $Q_1$ in Figure~\ref{fig:RPQ} yields the RPQ result pair $(v_0, v_9)$, obtained through the 4-hop path $(v_0 \rightarrow v_3 \rightarrow v_{12} \rightarrow v_{13} \rightarrow v_9)$.
Nevertheless, a DFS-based method would first explore a longer 6-hop path such as $(v_0 \rightarrow v_1 \rightarrow v_{10} \rightarrow v_{11} \rightarrow v_{12} \rightarrow v_{13} \rightarrow v_9)$.

Moreover, as the hop length increases, the amount of data that must be maintained in GPU and shared memory during traversal\,(e.g., slices, DFS stacks) also grows, making it impractical to set a sufficiently large hop bound.
Since the entire data graph often cannot fit into GPU memory, traversal is typically performed by partitioning the graph and loading only a subset~\cite{park2025cumatch, oh2025gflux, guo2020gpu, lai2022accelerating, zinn2016general}.
In practice, this constraint limits the hop bound to only a few dozen.
DFS-based traversal can easily exceed these bounds, leading to missed valid paths and ultimately incorrect results.
As shown in the experiments of Section~\ref{sec:experiment_RPQ}, BFS completed traversal within 5 hops, while the DFS-based hybrid approach explored up to 40 hops --- the maximum range allowed by GPU and shared memory --- but still failed to produce correct results.

\vspace*{0.1cm}
\textbf{Challenge 2\,(Limited parallelism due to visited set maintenance):}
As discussed in Section~\ref{sec:RPQ}, the automata-based approach maintains a visited set to avoid redundant exploration and ensure distinct vertex pairs.
The state-of-the-art CPU-based method Ring-RPQ~\cite{arroyuelo2024optimizing} also manages a visited set for each starting vertex using a bitmap.
In this case, the minimum memory usage per starting vertex is defined as $\frac{|V| \times |Q|}{8}$\,(in bytes), where $|V|$ is the number of vertices in the data graph and $|Q|$ is the number of states in the automaton.
For the LDBC SNB dataset with scale factor 10 and RPQ Q4\,($|V| = 35.5\,\text{M}, |Q| = 5$) used in Section~\ref{sec:experiment_RPQ}, the memory requirement per starting vertex is $(35.5\,\text{M} \times 5) / 8 \approx 22.2\,\text{MB}$.

On GPUs, it is common to assign each starting vertex to a separate thread block to exploit parallelism~\cite{park2025cumatch, oh2025gflux, bisson2017high}.
Since GPUs are designed to activate thousands of thread blocks simultaneously, thousands of starting vertices should ideally be processed in parallel.
However, GPU memory limitations make this difficult to achieve in practice.
For example, in the same graph mentioned earlier, if 20\,GB of GPU memory is allocated for the visited-set buffer, only $20\,\text{GB} / 22.2\,\text{MB} \approx 922$ starting vertices can be processed concurrently.
This limitation significantly reduces the effective utilization of available thread blocks and consequently leads to performance degradation.

\textbf{Challenge 3\,(Inability\,to\,evaluate\,CRPQs\,due to the explosion of RPQ results):}
CRPQs can be efficiently processed by first materializing the results of individual RPQ atoms and then combining them through the WCOJ-based CQ processing method proposed in \cite{park2025cumatch}.
However, since all-pairs RPQs enumerate every pair, their results can rapidly surpass GPU memory capacity.
For example, in Figure~\ref{fig:LGF}(a), the block $P_{11}$ pointed to by (3, 3, 2) in GridMap consists of slice $S_{11}$, which contains four edges forming a cycle over the vertex set $\{v_{10}, v_{11}, v_{12}, v_{13}\}$.
When evaluating $c^*$ on this slice, all vertices in the cycle become mutually reachable.
This result explosion also occurs on CPUs.
Even state-of-the-art CPU-based DBMSs Umbra~\cite{neumann2020umbra} and DuckDB~\cite{raasveldt2019duckdb}, which support CRPQs, can experience drastic increases in main-memory usage or even query failures during the materialization of RPQ atoms.
These systems follow the algebra-based approach and thus even the intermediate results must be materialized, leading to excessive memory consumption.

Recently, several studies have proposed methods for handling graph query results that exceed GPU memory capacity~\cite{lai2022accelerating, park2024infinel, he2008relational}.
These approaches typically enumerate large results rapidly and simply spill them into main memory.
In contrast, when the results need to be reused as input for subsequent operations\,(e.g., WCOJ), they must be converted into a GPU-aware graph format\,(e.g., LGF).
In particular, the parallel execution of GPU threads generates RPQ results in an irregular order~\cite{lai2022accelerating, park2024infinel}, so all results must be fully enumerated before conversion into a graph format.
However, converting on GPUs may exceed memory capacity, while converting on CPUs keeps the GPU idle until completion and reduces the benefits of acceleration.

\vspace{-0.1cm}
\section{GPU-Based Hop-Limited Level-Wise DFS} 
\label{sec:hop_limited_DFS}

In this section, we introduce the hop-limited level-wise DFS method, designed for efficient exploration of arbitrary path lengths in RPQs.
This method traverses the graph from a starting vertex in a fixed interval, referred to as a \textit{static-hop}.
It first explores all paths within the static-hop bound, and if additional extensions are required during traversal, the search range is progressively expanded by further static-hop intervals.
Hop-limited level-wise DFS can be described in two phases: (1) a \textbf{base phase}, which performs the initial exploration up to the static-hop bound, and (2) an \textbf{expansion phase}, which repeatedly extends the search range by static-hop intervals as needed.

\vspace{-0.1cm}
\subsection{Base Phase}
\label{sec:base_phase}

The base phase identifies the partitions\,(slices) to be explored within the static-hop bound using the automaton derived from the regular expression, and then performs traversal on the GPU.
We construct an \textit{RPQ traversal tree} that organizes candidate slices satisfying the regular expression into a tree structure, extending the concept proposed in~\cite{park2025cumatch}.
Starting from the initial state of the automaton and following its transitions, a \textit{candidate slice} is selected only if the following two conditions are satisfied:
\begin{itemize} [labelindent=0em,labelsep=0.3em,leftmargin=*,itemsep=0em]
\item The transition’s edge label matches the candidate slice’s label.
\item If a parent slice exists, it can connect to the candidate slice to form a path.
\end{itemize}
This connectivity condition is determined by checking the overlap of the source and destination vertex ID ranges\,(src-range, dst-range), which represent only actual vertices per slice and are precomputed during LGF construction.
Each node in the RPQ traversal tree stores its slice together with the automaton state reached through the transition, allowing us to check if the node is in a final state.

\begin{figure}[b]
\vspace*{-0.5cm}
    \centerline{\includegraphics[width=3.5in]{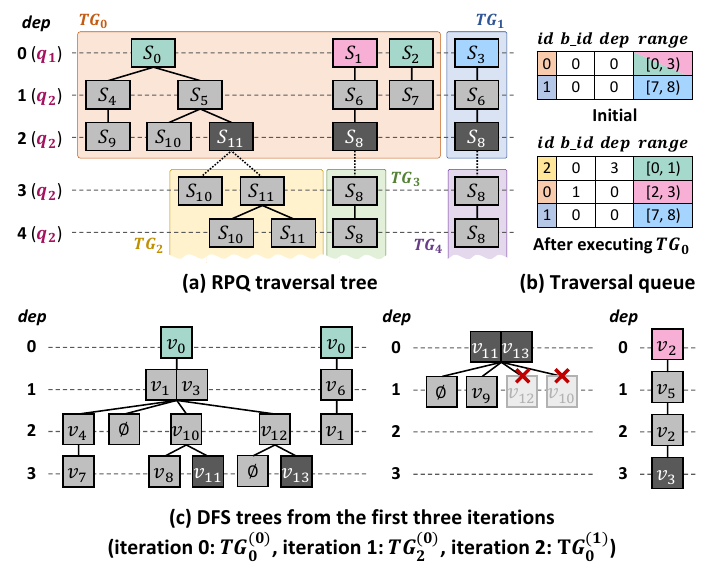}}
    \vspace*{-0.45cm}
    \caption{Example of hop-limited level-wise DFS.}
    \label{fig:hop_limited_DFS}
    \vspace*{-0.2cm}
\end{figure}

\textbf{Example:} Figure~\ref{fig:hop_limited_DFS}(a) shows the RPQ traversal tree for $Q_1$.
We assume the vertex ID of $v_i$ to be $i$, and use the plan in Figure~\ref{fig:AM_based}(a).
From the initial state $q_0$, the outgoing transition labeled $a$ selects slices $S_0$, $S_1$, $S_2$, and $S_3$ as the roots of the tree.
For $S_0$, the dst-range [1, 4) is checked for overlap with the src-ranges of slices labeled $b$.
Since the src-ranges of $S_4$ and $S_5$ are [1, 2) and [1, 4), respectively, these slices are considered candidates, while the others\,(e.g., $S_6$, $S_7$) are pruned.
Assuming a static-hop of 3, the tree construction terminates at depth 2.
The tree nodes at depth 0 reach automaton state $q_1$, while those at depths 1 and 2 reach state $q_2$.
The path through slices $S_0 \rightarrow S_4 \rightarrow S_9$\,(e.g., $v_0 \rightarrow v_1 \rightarrow v_4 \rightarrow v_7$) matches $abc$ and reaches the final state of the automaton, producing a valid RPQ result.

Since LGF is a 2D format partitioned by source and destination, root slices with the same block row ID can share starting vertices\,(e.g., both $S_0$ and $S_2$ include $v_0$).
To enable traversal at the granularity of starting vertices, such subtrees are grouped into a \textit{Traversal Group}\,(TG), which serves as the basic processing unit of traversal.
For example, in Figure~\ref{fig:hop_limited_DFS}(a), two TGs are generated in the base phase, since slices $S_0$, $S_1$, and $S_2$ have block row ID 0, while $S_3$ has block row ID 2.
TG IDs are assigned sequentially, and the TG with ID $i$ is denoted $TG_i$.

After the slices that compose a TG are loaded into an \textit{input buffer} in GPU memory, each TG is processed on the GPU in two steps: (1) \textit{determining the set of starting vertices} and (2) \textit{performing path exploration}.
As we pointed out in Section~\ref{sec:challenges}, processing all starting vertices at once is infeasible in all-pairs RPQs due to GPU memory constraints.
Therefore, in the first step, the starting vertices are divided into batches of fixed size, referred to as the \textit{batch size}, determined by GPU occupancy\,(e.g., 4,096 concurrent starting vertices).
In detail, the starting vertices for each batch are determined by performing a \textit{k-way merge}~\cite{wilkes1974art} over the source vertex arrays of the TG’s root slices.
In the second step, path exploration is performed on the TG using the DFS method proposed in~\cite{park2025cumatch}.
Traversal begins at the root slice with each starting vertex.
Paths are then extended one hop along the edges in each slice, and the newly reached vertices are recursively explored through the child slices.
Each batch in $TG_i$ is assigned a sequential \textit{batch ID} starting from 0, and the batch with ID $j$ is denoted $TG_i^{(j)}$. 
A TG is fully processed once the starting vertices of all its root slices have been completely explored.

\textbf{Example:} We consider $TG_0$ in Figure~\ref{fig:hop_limited_DFS}(a).
Assume that the eleven slices comprising $TG_0$ have been fully loaded into GPU memory.
The source vertex sets of $S_0$, $S_1$, and $S_2$ are $\{v_0\}$, $\{v_2\}$, and $\{v_0\}$, respectively.
When the batch size is set to 1, batch 0 selects $\{v_0\}$ as the starting vertex set.
The left side of Figure~\ref{fig:hop_limited_DFS}(c) shows the DFS tree of $TG_0^{(0)}$, which depicts the exploration order of the path exploration process.
Since $S_1$ does not contain any edges starting from $v_0$, it is excluded, and exploration begins from $S_0$ and $S_2$.
From $v_0$ through $S_0$, vertices $\{v_1, v_3\}$ are reached, which then leads to exploration of $S_4$ and $S_5$.
Following DFS order, exploration first proceeds through $S_4$, yielding the 3-hop path $(v_0 \rightarrow v_1 \rightarrow v_4 \rightarrow v_7)$ via $S_9$.
Next, from $S_5$, traversal continues through $S_{10}$ and $S_{11}$, reaching $\{v_8\}$ and $\{v_{11}, v_{13}\}$, respectively.
Once the exploration of $TG_0^{(0)}$ is completed, batch 1 explores paths starting from $\{v_2\}$\,(the right side of Figure~\ref{fig:hop_limited_DFS}(c)).
In this example, $TG_0$ is fully processed after two batches.
$TG_2^{(0)}$ will be explained later.

\vspace{-0.1cm}
\subsection{Expansion Phase}
\label{sec:expansion_phase}

Whenever each batch of a TG in the base phase completes and traversal can proceed beyond the static-hop boundary, the expansion phase is triggered.
The expansion phase extends the RPQ traversal tree by the length of the static-hop from the leaf nodes of the parent TG at the static-hop boundary and continues the exploration.
Each sub-tree extended from a single leaf node is defined as a new TG.
Such a TG is referred to as an \textit{expansion-TG}, whereas a TG generated in the base phase is called a \textit{base-TG}.
During the processing of an expansion-TG, the same expansion procedure can be recursively applied whenever additional extension is required.
The base-TG determines its own set of starting vertices before traversal, whereas the expansion-TG inherits the set from its parent and executes only the second step\,(i.e., the path exploration step).
At this step, traversal continues from the \textit{checkpoint set} of each starting vertex, which contains the vertices reached at the static-hop boundary.
We define the \textit{checkpoint set} in Definition~\ref{definition:checkpoint_vertex_set}.

\vspace*{-0.05cm}
\begin{definition} 
\label{definition:checkpoint_vertex_set} 
\textbf{Checkpoint set}:
We define the \textit{checkpoint set} for a start vertex $s$ as the set of vertices reached at the static-hop boundary from the slices corresponding to leaf nodes of the parent TG.
This set serves as the \textit{exploration frontier} for $s$ in the expansion phase.
\end{definition}
\vspace{-0.05cm}

\textbf{Example:} In Figures~\ref{fig:hop_limited_DFS}(a) and (c), the processing of $TG_0^{(0)}$ reaches two paths, $(v_0 \rightarrow v_1 \rightarrow v_{10} \rightarrow v_{11})$ and $(v_0 \rightarrow v_3 \rightarrow v_{12} \rightarrow v_{13})$, through $S_{11}$, which corresponds to a leaf node located at the static-hop boundary.
In this case, we obtain the checkpoint set $\{v_{11}, v_{13}\}$ from the starting vertex $v_0$.
The expansion phase is performed when candidate slices connected to the checkpoint set, serving as the frontier, are available.
In detail, expansion at $S_{11}$ occurs when the automaton contains a transition labeled $c$\,(e.g., a self-loop on $q_2$ with label $c$) and, at the same time, the vertex ID range of the checkpoint set $[11, 14)$ overlaps with the src-range of slices in the data graph\,(e.g., $S_{10}$ and $S_{11}$).
The sub-tree derived from $S_{11}$ is grouped into a single expansion-TG, denoted $TG_2$.
The middle part of Figure~\ref{fig:hop_limited_DFS}(c) shows the DFS tree of $TG_2^{(0)}$.
Here, traversal resumes from the checkpoint set $\{v_{11}, v_{13}\}$, and an additional path $(v_{13} \rightarrow v_9)$ is discovered through $S_{10}$.
This yields the RPQ result $(v_0, v_9)$.
Although traversal from $S_{11}$ can reach $v_{12}$ and $v_{10}$, these are paths already visited at depth 2 of $TG_0^{(0)}$ and are therefore skipped through the visited set management to be described in Section~\ref{sec:segment_pooling}.
Slices such as $S_9$ and $S_{10}$ yield 3-hop paths\,(e.g., $v_0 \rightarrow v_1 \rightarrow v_4 \rightarrow v_7$), but since no candidate slices are connected to these leaf nodes, no expansion phase is triggered.

Hop-limited level-wise DFS dynamically generates base-TGs and expansion-TGs during traversal, and their execution order is managed through a priority queue called the \textit{traversal queue}. 
Each queue record stores the \textit{TG ID}, \textit{batch ID}, \textit{depth}, and \textit{start-vertex ID range}.
TGs with greater depth have higher priority, and ties are broken by TG ID.
When a TG is executed, new expansion-TGs are inserted into the queue, and the record of the original TG is updated to the next batch until all its starting vertices are processed.
We call one such \textit{dequeue-execution-enqueue} cycle an \textit{iteration}, and traversal terminates when the queue becomes empty.

\textbf{Example:} 
Figure~\ref{fig:hop_limited_DFS}(b) shows how traversal progresses through the traversal queue.
For brevity, each record is denoted as $id$, $b\_id$, $dep$, and $range$.
Records placed higher in the queue represent those with higher priority.
In the initial state, the records of base-TGs $TG_0$ and $TG_1$ are inserted into the queue.
The execution begins with $TG_0^{(0)}$, and once traversal with the starting vertex set $\{v_0\}$ is completed, the expansion phase generates $TG_2$.
Here, the depth of $TG_2$ is 3, and its start-vertex ID range is inherited from the starting vertex set of $TG_0^{(0)}$, resulting in [0, 1).
Subsequently, the batch ID of $TG_0$ is incremented to 1, and since the processing of $v_0$ has been completed, its start vertex range is updated to [2, 3).
The entire execution consists of six iterations, following the order $TG_0^{(0)} \rightarrow TG_2^{(0)} \rightarrow TG_0^{(1)} \rightarrow TG_3^{(0)} \rightarrow TG_1^{(0)} \rightarrow TG_4^{(0)}$.

\vspace*{0.05cm}
Hop-limited level-wise DFS differs from naive DFS by following depth-first exploration while collectively processing all paths within each static-hop boundary.
This design effectively \textit{incorporates the shortest-path-first property of BFS into arbitrary-path length queries,} reducing redundant exploration.
For example, when naive DFS is executed in vertex ID order, the RPQ result $(v_0, v_9)$ can only be reached via a \textbf{6-hop} path such as $(v_0 \rightarrow v_1 \rightarrow v_{10} \rightarrow v_{11} \rightarrow v_{12} \rightarrow v_{13} \rightarrow v_9)$.
In contrast, as shown in Figure~\ref{fig:hop_limited_DFS}(c), $TG_0^{(0)}$ first discovers the path $(v_0 \rightarrow v_3 \rightarrow v_{12} \rightarrow v_{13})$, after which $TG_2^{(0)}$ extends it with $(v_{13} \rightarrow v_9)$, reaching the same result in only \textbf{4 hops}.

\vspace{-0.1cm}
\section{GPU-Based On-Demand Segment Pooling}
\label{sec:segment_pooling}

In this section, we explain the on-demand segment pooling technique, designed for the efficient management of visited sets within GPU memory.
During each iteration, when traversing the batches of a TG, it is necessary to maintain the visited sets for up to the maximum batch size of starting vertices in order to detect duplicate paths and prevent redundant exploration\,(e.g., $v_{12}$ and $v_{10}$ in the middle part of Figure~\ref{fig:hop_limited_DFS}(c)).
Thanks to the RPQ traversal tree and the LGF format, the slices to be examined during traversal are predetermined, and the maximum size of the reachable vertex set for each slice is bounded by the \textit{block-level width}\,(e.g., four in Figure~\ref{fig:LGF}(a)).
Thus, rather than maintaining a global visited set for the entire graph, only the visited sets required for traversal are managed in units called \textit{segments}.

A segment is a logical unit formed by partitioning a pre-allocated \textit{segment buffer} in GPU memory into bitmap regions of block-level width.
Each segment is assigned a sequential ID starting from 0.
Free segments are managed in a \textit{segment pool}, and each TG node\,(e.g., eleven TG nodes for $TG_0$ in Figure~\ref{fig:hop_limited_DFS}(a)) is assigned the required segment IDs from the pool at the beginning of the path exploration step.
Once the traversal of the current TG completes and the segments are no longer used, they are returned to the pool.
Segments are not limited to visited set management but may serve multiple purposes, classified into three principal types:
(1) \textbf{Visited segments}, which record previously traversed paths to prevent redundant exploration, (2) \textbf{Checkpoint segments}, which preserve vertices reached at the static-hop boundary\,(i.e., the checkpoint set) so that expansion-TGs may continue traversal,
and (3) \textbf{Bridge segments}, which maintain traversal continuity when a TG is too large to fit in GPU memory, allowing execution at the sub-TG level.

\vspace{-0.25cm}
\subsection{Visited Segment}
\label{sec:visited_segment}

Each segment is uniquely identified by a search context defined as (\textit{starting vertex ID}, \textit{automaton state}, \textit{column ID}).
Segments are assigned based on this key, and nodes with the same key share the same segment.
Here, the column ID refers to the block column ID of the block to which a slice belongs. 
For example, in Figure~\ref{fig:LGF}, both $S_9$ and $S_{10}$ have column ID 2.
This mapping is managed through the \textit{segment pooling table}, and if no segment exists for a given key, a new one is allocated from the segment pool.
For single-source RPQs, each TG node is assigned exactly one visited segment, whereas for all-pairs RPQs, each node is assigned a number of visited segments equal to the batch size.

\begin{figure}[b]
\vspace*{-0.7cm}
    \centerline{\includegraphics[width=3.4in, trim=4 0 0 0, clip]{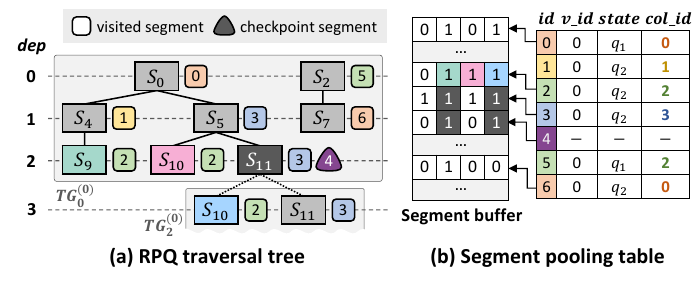}}
    \vspace*{-0.4cm}
    \caption{Example of on-demand segment pooling method.}
    \label{fig:segment_pooling}
    \vspace*{-0.2cm}
\end{figure}

\textbf{Example:}
Figure~\ref{fig:segment_pooling} shows how $TG_0^{(0)}$ in Figure~\ref{fig:hop_limited_DFS} acquires visited segments using on-demand segment pooling.
The starting vertex set of $TG_0^{(0)}$ is $\{v_0\}$.
Among the sub-trees of $TG_0^{(0)}$, the one rooted at $S_1$ is omitted because it does not contain $v_0$.
The rounded rectangles indicate the segment IDs mapped as visited segments to each node.
Figure~\ref{fig:segment_pooling}(b) shows the segment buffer and the segment pooling table, where the buffer holds several segments represented as 4-bit bitmaps.
For brevity, the segment ID is denoted as $id$, and the key as $(v\_id, state, col\_id)$.
Among the eight TG nodes, $S_9$ and $S_{10}$ have the same segment ID, as do $S_5$ and $S_{11}$\,(e.g., 2 and 3, respectively).
During the path exploration step, traversal reaches $\{v_7\}$ from $S_9$ and $\{v_8\}$ from $S_{10}$.
This sets the second and third bits of the bitmap for segment ID 2.
In contrast, $S_0$ and $S_7$ share the same column ID\,(e.g., 0) but correspond to different automaton states\,(e.g., $q_1$ and $q_2$), and thus map to different segment IDs.

A visited segment is released once the traversal of a TG and all of its derived expansion-TGs is complete.
For example, in Figure~\ref{fig:segment_pooling}, after $TG_0^{(0)}$ completes, the visited segments are still retained for the exploration of the derived $TG_2^{(0)}$.
When $TG_2^{(0)}$ completes, the related visited segments\,(e.g., six in total) are released.
Since the traversal queue prioritizes deeper TGs, visited segments are returned quickly instead of being retained for extended periods.

\vspace{-0.1cm}
\subsection{Checkpoint Segment}
\label{sec:checkpoint_segment}

A checkpoint segment stores the checkpoint set in bitmap form, enabling exploration to continue within an expansion-TG.
Each leaf node at the static-hop boundary allocates checkpoint segments.
In a single-source RPQ, each leaf node is assigned exactly one checkpoint segment, whereas in an all-pairs RPQ, each leaf is assigned a number of checkpoint segments equal to the batch size.
A checkpoint segment is released once its expansion-TG completes.

\textbf{Example:} In Figure~\ref{fig:segment_pooling}(a), the leaf node $S_{11}$ is assigned a checkpoint segment with ID 4.
During traversal, the second and fourth bits of this segment are set to represent the checkpoint set $\{v_{11}, v_{13}\}$.
This set then serves as the traversal frontier for the expansion-TG, $TG_2$.
When $TG_2$ completes, the checkpoint segment\,(ID = 4) is released.
If there are no candidate slices for the next hop, no checkpoint segment is allocated\,(e.g., $S_9$ and $S_{10}$).

\begin{figure}[b]
\vspace*{-0.4cm}
    \centerline{\includegraphics[width=3.3in]{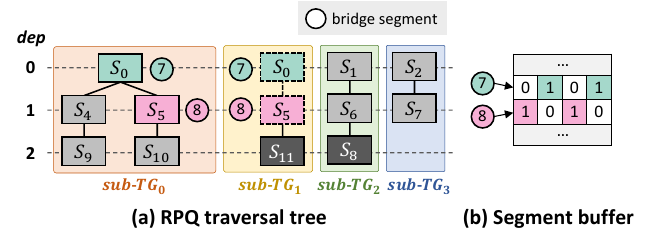}}
    \vspace*{-0.4cm}
    \caption{Example of sub-TG partitioning.}
    \label{fig:continuous_segment}
    \vspace*{-0.2cm}
\end{figure}

\vspace{-0.1cm}
\subsection{Bridge Segment}
\label{sec:bridge_segment}

When the data graph is large or the exploration range broad, processing a single TG in one execution may be difficult.
For example, the slices of the TG may not all fit into the input buffer, or even when all segments are utilized, execution may still be infeasible.
To address this issue, cuRPQ applies \textit{sub-TG partitioning}, which divides a TG into smaller units called \textit{sub-TGs}.
A sub-TG is defined along a \textit{root-leaf tree path} of the TG, where the consecutive nodes along the tree path to each leaf are grouped into a sub-TG.
When multiple leaves share ancestor nodes, those nodes are duplicated across the corresponding sub-TGs.
During this process, the input buffer usage and the number of segments are cumulatively estimated along each tree path.
If execution becomes infeasible, the tree path explored so far is finalized as a sub-TG, after which the procedure proceeds with the remaining nodes.
For example, Figure~\ref{fig:continuous_segment}(a) shows $TG_0$ partitioned into four sub-TGs, $sub$-$TG_{0..3}$. 
In this case, the tree paths $\{S_0, S_4, S_9\}$ and $\{S_0, S_5, S_{10}\}$ belong to $sub$-$TG_0$, while the path $\{S_0, S_5, S_{11}\}$ forms a separate $sub$-$TG_1$.

When a TG is divided into multiple sub-TGs, \textbf{two major issues} arise.
\textbf{First,} \textit{because the same starting vertex may appear in multiple sub-TGs, its visited segments must be retained until all relevant sub-TGs have completed.}
To avoid this, we extend the priority policy of the traversal queue so that sub-TGs are executed in ascending order of their start-vertex ID ranges, ensuring that each starting vertex is processed sequentially.
We release the visited segments only after the vertex has been fully processed across all sub-TGs.
For example, in Figure~\ref{fig:continuous_segment}(a), the start-vertex ID ranges of $sub$-$TG_{0..3}$ are [0, 1), [0, 1), [2, 3), and [0, 1), respectively. 
Thus, $sub$-$TG_0$, $sub$-$TG_1$, and $sub$-$TG_3$ are processed in order, after which the visited segments for $v_0$ are released, followed by the execution of $sub$-$TG_2$.

\textbf{Second,} \textit{when consecutive sub-TGs share common nodes at their boundary, path exploration may be prematurely blocked, since those nodes could have already been marked as visited in an earlier sub-TG.}
We refer to these common nodes as the \textit{cut set}.
For example, the cut set between $sub$-$TG_0$ and $sub$-$TG_1$ is $\{S_0, S_5\}$.
If $\{v_1, v_3\}$ is first reached through $S_0$ in $sub$-$TG_0$ and marked as visited, then in $sub$-$TG_1$, the path through slices $S_0 \rightarrow S_5 \rightarrow S_{11}$ can no longer be explored.

We address this issue by storing the reachable vertices of the cut-set nodes in bridge segments and passing them to the subsequent sub-TG.
This ensures that when the same slice is revisited, exploration proceeds only through the permitted vertices recorded in the bridge segment, regardless of the visited segment.
For example, in Figure~\ref{fig:continuous_segment}, when traversing $sub$-$TG_0$, vertices $\{v_1, v_3\}$ from $S_0$ and $\{v_{10}, v_{12}\}$ from $S_5$ are recorded in segments 7 and 8, respectively.
When $sub$-$TG_1$ is executed, the bridge segments are passed from $sub$-$TG_0$, and traversal continues by following only the edges leading to vertices whose bits are set, regardless of the visited set.
In a single-source RPQ, each cut-set node is assigned one bridge segment, whereas in an all-pairs RPQ, it is assigned a number of bridge segments equal to the batch size.
Once the subsequent sub-TG completes, the corresponding bridge segments are released.

\vspace{-0.1cm}
\section{Concurrent Exploration-Materialization}
\label{sec:intermediate_data_control}

In this section, we explain the concurrent exploration-materialization strategy in cuRPQ, designed to efficiently manage large RPQ results.
Section~\ref{sec:incremental_materialization} introduces the \textit{batch-incremental materialization} method, which enables concurrent GPU-side exploration and CPU-side materialization.
Section~\ref{sec:CRPQ_and_Plans} describes how this method extends to CRPQ processing and execution plan strategies.

\vspace{-0.1cm}
\subsection{Batch-Incremental Materialization}
\label{sec:incremental_materialization}

To store large RPQ results efficiently on the GPU and enable their reuse in subsequent operations\,(e.g., CRPQ or RPQ), cuRPQ materializes the results in the LGF format.
The result of a single RPQ forms a grid, where each node in the RPQ traversal tree corresponds to a block.
A TG corresponds to the blocks within a single row of this grid, all of which share the same block row ID.
For example, in Figure~\ref{fig:hop_limited_DFS}(a), $S_4$ of $TG_0$ corresponds to coordinate $(0, 1)$, defined by the block row ID of its root node and the block column ID of the slice, while both $S_9$ and $S_{10}$ map to $(0, 2)$.
A straightforward approach would be to allocate a GPU memory buffer for each block and store the corresponding RPQ results directly.
However, block-level result sizes are difficult to predict in advance.
This leads to excessive space allocation, and large outputs can easily exceed GPU memory capacity, making execution infeasible.

cuRPQ adopts an approach where results from all blocks are aggregated into a \textit{unified RPQ results buffer}\,(UR buffer).
Following the technique proposed in~\cite{park2024infinel}, the UR buffer is divided into small chunks, and each thread writes results sequentially into its assigned chunk.
When a chunk becomes full, a new chunk is allocated.
If no more chunks are available, the kernel suspends execution, transfers the accumulated results from the UR buffer to main memory, and then resumes kernel execution.
This strategy keeps GPU memory usage stable.
However, the drawback remains that RPQ results must still be materialized in bulk, leading to high reconstruction overhead.
Furthermore, ensuring that each slice remains below a threshold $\theta$\,(e.g., block $P'_{1}$ in Figure~\ref{fig:LGF}(b)) requires repeated recursive quadrant partitioning, which often becomes a major bottleneck.

We explain a \textit{Batch-Incremental Materialization}\,(BIM) method that allows the CPU to progressively materialize RPQ results during query execution.
The key idea is to accumulate results block by block during each UR buffer transfer and to progressively materialize completed slices on the CPU during query execution.
Since both TGs and sub-TGs are processed in batches ordered by starting vertex IDs, the results within each block are also filled in this order.
Thus, once the exploration for a certain vertex range has finished, the corresponding slices can be safely materialized.
This BIM process consists of three steps.

\begin{figure}[b]
\vspace*{-0.3cm}
    \centerline{\includegraphics[width=3.4in]{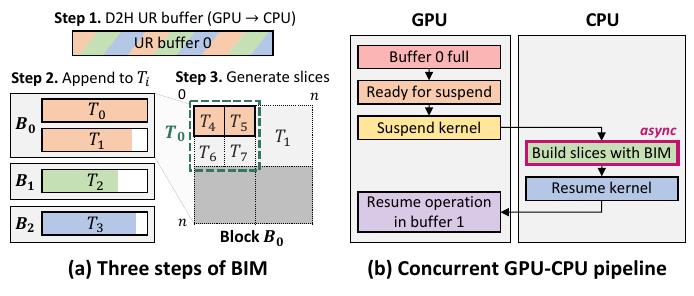}}
    \vspace*{-0.4cm}
    \caption{Example of concurrent exploration-materialization.}
    \label{fig:RPQ_result_control}
\end{figure}

\begin{itemize}[labelindent=0em,labelsep=0.3em,leftmargin=*,itemsep=0em]

\item \textbf{Step~1:} RPQ results accumulated in the UR buffer are transferred from the GPU to main memory via device-to-host\,(D2H) transfer.

\item \textbf{Step~2:} The CPU scans the transferred results, determines the block to which each result belongs, and stores them in the corresponding \textit{temporary slice buffer}\,($T_i$).
Each $T_i$ is an intermediate buffer with a capacity of $\theta$, used to accumulate results before slice materialization.
Initially, one $T_i$ is allocated per block, but when a buffer becomes full, the slice is partitioned into quadrants and up to four new $T_i$ buffers are allocated.

\item \textbf{Step~3:} When the exploration of a slice’s starting vertex range completes, its corresponding $T_i$ is materialized as a slice.

\end{itemize}

Figure~\ref{fig:RPQ_result_control}(a) shows an example of BIM.
At this point, results are being produced for three blocks, and four temporary buffers $T_{0..3}$ are maintained in main memory. 
Assuming the block width and height are $n$, $T_0$ covers the range $[0, n/2) \times [0, n/2)$ and $T_1$ covers $[0, n/2) \times [n/2, n)$. 
By contrast, $T_2$ and $T_3$ have not been partitioned yet and still cover the entire range $[0, n) \times [0, n)$. 
During Step~2, if $T_0$ becomes full, it is replaced by four new buffers $T_4$-$T_7$ created through quadrant partitioning.
The results stored in $T_0$ are redistributed among these new buffers based on their ranges, and $T_0$ is released.
During Step~3, once the TG\,(or its sub-TGs) associated with block $B_0$ has completed exploration of the starting vertex range $[0, n/4)$, the corresponding buffers $T_4$ and $T_5$ are materialized as slices.

This approach can be extended to support asynchronous CPU-side BIM, which overlaps with the RPQ kernel through GPU streams.
Figure~\ref{fig:RPQ_result_control}(b) shows a GPU-CPU pipeline that overlaps query execution with materialization by alternating between two UR buffers.
When UR buffer~0 becomes full, the kernel pauses briefly and signals the CPU to perform the BIM step asynchronously, starting with a D2H transfer\,(Step 1).
Meanwhile, the GPU resumes exploration using UR buffer~1.
By alternating between the two buffers, query execution and materialization are overlapped.

\vspace{-0.1cm}
\subsection{Processing CRPQs with Execution Plans}
\label{sec:CRPQ_and_Plans}

After materializing the results of RPQ atoms, we process a CRPQ by combining them using the WCOJ-based CQ method proposed in~\cite{park2025cumatch}.
However, depending on the matching order in WCOJ~\cite{park2025cumatch, lai2022accelerating, zinn2016general}, the exploration direction of an RPQ atom may differ from the direction required in the join.
For example, even if exploration is performed and materialized in the out-edge direction\,(i.e., out-edge slice), WCOJ may require an in-edge slice instead.
To address this, the CRPQ plan uses the matching order to determine the required direction for each RPQ atom in advance and materializes results only in that direction.
If the opposite direction is required, we perform a \textit{slice transpose}, denoting the corresponding automaton $A$ as ${A}^T$ in the plan.

\begin{figure}[htbp]
\vspace*{-0.2cm}
    \centerline{\includegraphics[width=3.4in, trim=0 0 4 10, clip]{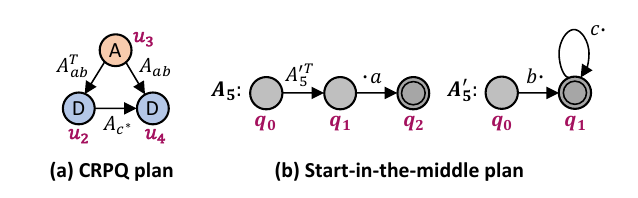}}
    \vspace*{-0.45cm}
    \caption{Example of execution plan strategies.}
    \label{fig:CRPQ_and_execution_plans}
    \vspace*{-0.2cm}
\end{figure}

Figure~\ref{fig:CRPQ_and_execution_plans}(a) shows the CRPQ plan for $Q_2$ in Figure~\ref{fig:RPQ}. 
Each edge is annotated with the automaton of its corresponding RPQ atom, indicating that their results will be joined using WCOJ.
Suppose the automata for $x \xrightarrow{ab} y$ and $x \xrightarrow{c^*} y$ are $A_{ab}$ and $A_{c^*}$, respectively, and assume the matching order $u_2 \prec u_3 \prec u_4$.
In this case, $A_{c^*}$ requires only out-edge slices, whereas $A_{ab}$ requires both out-edge and in-edge slices for WCOJ.
The key point is that while $A_{ab}$ requires results in both directions, exploration is done only once, with materialization producing both out-edge and in-edge slices.

In addition, cuRPQ can support diverse execution plan strategies such as WavePlan~\cite{yakovets2016query}.
For example, the \textit{reverse} plan\,(e.g., $A_1$ in Figure~\ref{fig:WavePlan}) can be handled by using in-edge slices in the LGF format\,($GridMap_{in}$).
The \textit{loop-cache} plan\,(e.g., $A_2$) leverages cuRPQ’s materialization strategy, where partial query results\,(e.g., $A'_2$) are first materialized as slices and then reused in subsequent RPQ exploration to extend paths.
In contrast, the \textit{start-in-the-middle} plan\,(e.g., $A_3$ and $A_4$) begins exploration from the middle of the path, so result pairs cannot be confirmed in order of start vertices as in forward exploration.
Such plans must first enumerate all result pairs before materialization, and therefore BIM cannot be applied directly.
To execute these plans under limited GPU memory, cuRPQ preserves the start-in-the-middle strategy while leveraging slice transpose.
In this approach, one sub-RPQ result is materialized in a single direction and then transposed into the opposite direction for joining.
For example, Figure~\ref{fig:CRPQ_and_execution_plans}(b) shows cuRPQ’s plan for $A_3$, where the out-edge paths satisfying $bc^*$\,(e.g., $A'_5$) are first computed and then transposed into in-edge slices to join with $a$\,(e.g., $A_5$).

\section{Architecture of cuRPQ}
\label{sec:architecture}

\begin{figure}[b]
\vspace*{-0.1cm}
    \centerline{\includegraphics[width=3.2in]{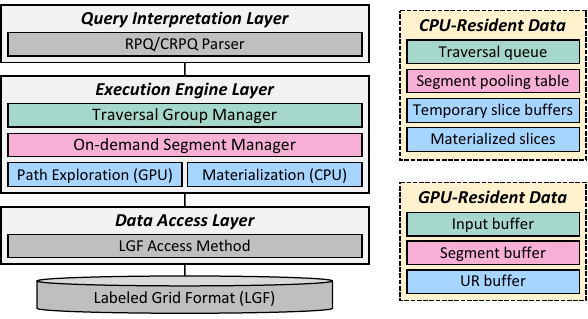}}
    \vspace*{-0.15cm}
    \caption{Architecture of cuRPQ and its core metadata structures.}
    \label{fig:architecture}
    \vspace*{-0.2cm}
\end{figure}

Figure~\ref{fig:architecture} provides an overview of the cuRPQ architecture, which consists of three main layers: the \emph{query interpretation layer}, the \emph{execution engine layer}, and the \emph{data access layer}.
Given an RPQ or CRPQ as input, the query interpretation layer parses the query and translates it into a corresponding automata-based execution plan.
The execution engine layer then performs RPQ traversal based on this automaton.
Specifically, the traversal group manager constructs the RPQ traversal tree and manages TGs via the traversal queue\,(Section~\ref{sec:hop_limited_DFS}).
For each TG, the on-demand segment manager allocates the required segments during traversal\,(Section~\ref{sec:segment_pooling}).
Path exploration is executed on the GPU, while result materialization is performed concurrently on the CPU\,(Section~\ref{sec:intermediate_data_control}).
At the data access layer, graph data stored on disk in the LGF format is accessed on demand during execution.

Algorithm~\ref{alg:cuRPQ_execution} presents the overall workflow of the execution engine layer.
cuRPQ processes each RPQ atom in the execution plan sequentially\,(Line~\ref{alg:cuRPQ:for1}).
For each atom, it first executes the base phase\,(Lines~\ref{alg:cuRPQ:base-phase1}-\ref{alg:cuRPQ:base-phase2}).
In each iteration, the execution engine performs traversal for a TG selected from the traversal queue\,(Lines~\ref{alg:cuRPQ:while1}-\ref{alg:cuRPQ:segment2}) and extends the TG via the expansion phase when necessary to enable further traversal\,(Lines~\ref{alg:cuRPQ:expansion-phase1}-\ref{alg:cuRPQ:expansion-phase2}).
During the execution of each TG, the engine loads the required slices, allocates the necessary segments, and launches the path exploration kernel on the GPU\,(Lines~\ref{alg:cuRPQ:load}-\ref{alg:cuRPQ:exploration}).
When the UR buffer reaches its capacity, the execution engine asynchronously materializes slices in the UR buffer on the CPU and then resumes GPU kernel execution, thereby overlapping exploration and materialization\,(Lines~\ref{alg:cuRPQ:materialize1}-\ref{alg:cuRPQ:materialize3}).
For CRPQs, once all RPQ atoms have been processed, the system performs WCOJ-based conjunctive query processing over the materialized slices\,(Lines~\ref{alg:cuRPQ:CRPQ1}-\ref{alg:cuRPQ:CRPQ2}).

\begin{algorithm}[t]
\SetAlgoLined
\SetArgSty{textnormal}

\KwIn{RPQ/CRPQ execution plan $P$}

\vspace{0.1cm}
\ForEach{RPQ atom $a$ in $P$}{
\label{alg:cuRPQ:for1}
    Generate base-TGs for $a$\;
\label{alg:cuRPQ:base-phase1}
    Enqueue base-TGs into traversal queue\;
\label{alg:cuRPQ:base-phase2}
    
    \While{traversal queue is not empty}{
\label{alg:cuRPQ:while1}
        $tg \leftarrow$ Dequeue a TG from traversal queue\;
\label{alg:cuRPQ:dequeue}
        Load slices of $tg$ into GPU input buffer\;
\label{alg:cuRPQ:load}
        Allocate required segments for $tg$\;
\label{alg:cuRPQ:segment1}
        Launch the path exploration kernel for $tg$\;
\label{alg:cuRPQ:exploration}

        \While{UR buffer is full}{
\label{alg:cuRPQ:materialize1}
            Asynchronously materialize RPQ results\;    
\label{alg:cuRPQ:materialize2}
            Resume the path exploration kernel for $tg$\;
\label{alg:cuRPQ:materialize3}
        }

        \If{expansion phase is triggered for $tg$}{
\label{alg:cuRPQ:expansion-phase1}
            Enqueue expansion-TGs into traversal queue\;
\label{alg:cuRPQ:expansion-phase2}
        }
        
        Release allocated segments of $tg$\;
\label{alg:cuRPQ:segment2}
    }
}

\If{$P$ is CRPQ}{
\label{alg:cuRPQ:CRPQ1}
    Execute a WCOJ-based CQ kernel on materialized slices\;
\label{alg:cuRPQ:CRPQ2}
}

\caption{\textsc{Execution Workflow in cuRPQ}}
\label{alg:cuRPQ_execution}
\end{algorithm}

The path exploration kernel utilizes all available GPU thread blocks.
Following widely adopted strategies in GPU-based graph analytics~\cite{oh2025gflux, bisson2017high, park2025cumatch}, each starting vertex in a TG is mapped to a separate thread block to exploit parallelism.
Within each thread block, threads process the adjacency list of the slice corresponding to the current TG node during hop-limited level-wise DFS.
Destination vertices are accessed in a coalesced manner, enabling efficient warp-level memory access.
Due to hop-limited traversal, each TG forms a tree structure that is small enough to reside in shared memory, thereby reducing random global memory accesses during traversal.

Figure~\ref{fig:architecture} also shows the core metadata structures maintained during execution.
The traversal queue prioritizes deeper TGs, which prevents the queue from continuously growing with shallower TGs and keeps the queue size small.
For the LDBC SNB dataset with scale factor 10 and RPQ Q4 used in Section~\ref{sec:experiment_RPQ}, the traversal queue maintains at most 34 entries, resulting in only a few kilobytes of overhead.
The segment pooling table size is bounded by the number of pre-allocated segments and remains within a few megabytes.
In addition, the input buffer, segment buffer, and UR buffer are allocated with fixed sizes on the GPU, avoiding additional memory allocation during execution.

\vspace{-0.1cm}
\section{Performance Evaluation}
\label{sec:performance_evaluation}



\vspace{-0.05cm}
\subsection{Experimental Setup}
\label{sec:setup}

\textbf{RPQ datasets and queries:}
We adopt datasets and queries widely used in prior studies on RPQs in streaming graph environments~\cite{pacaci2020regular, gou2024lm, pacaci2022evaluating}.
The ten RPQ queries used in our evaluation are summarized in Table~\ref{tab:queries}.
Among them, nine are recursive queries that frequently appear in real-world applications~\cite{bonifati2019navigating, bonifati2020analytical}, while Q4 is a non-recursive query that we slightly modified for use as a baseline.
These recursive query patterns have been reported to cover more than 99\% of all recursive RPQs observed in Wikidata query logs~\cite{bonifati2019navigating}.
For the data graphs, we use both synthetic and real-world datasets.
As the synthetic dataset, we use the \textbf{LDBC SNB}~\cite{szarnyas2022ldbc}, which simulates real-world interactions in social networks. 
This dataset contains 12 types of vertex labels and 15 types of edge labels.
We use two \textit{scale factors}\,(SF): SF=1\,(4.0\,M vertices and 22.4\,M edges) and SF=10\,(35.5\,M vertices and 219.4\,M edges). 
Table~\ref{tab:result} shows the number of results for each query.
As the LDBC SNB dataset contains only two recursive edges\,(e.g., \textit{ReplyOf} and \textit{Knows}), some recursive queries such as Q8, Q9, and Q10 cannot be meaningfully represented; therefore, their result counts are omitted from Table~\ref{tab:result}, as in~\cite{pacaci2020regular, gou2024lm}.

As the real-world dataset, we use \textbf{StackOverflow}~\cite{snapnets}, which contains 3 types of vertex labels and 3 types of edge labels, spanning 7 years and 7 months. 
Based on timestamps, we divide it into two spans: Span=6M\,(22.1\,K vertices and 433.9\,K edges) and Span=1Y\,(46.6\,K vertices and 1.3\,M edges). 
All queries return the number of result pairs satisfying the given regular expression.

\begin{table}[t]
  \vspace*{-0.1cm}
  \caption{RPQ queries used in the experiments.}
  \vspace*{-0.3cm}
  \label{tab:queries}
  \begin{tabular}{c|c||c|c}
    \toprule
    \textbf{Notation} & \textbf{\parbox{1.5cm}{\centering Query}} & \textbf{Notation} & \textbf{\parbox{1.3cm}{\centering Query}} \\
    \midrule
    Q1 & $a^*$ & Q6 & $ab^*c$ \\
    Q2 & $a?b^*$ & Q7 & $(a_1 + a_2 + ... + a_k)b^*$ \\
    Q3 & $ab^*$ & Q8 & $a^*b^*$ \\
    Q4 & $abcd$ & Q9 & $ab^*c^*$ \\
    Q5 & $abc^*$ & Q10 & $(a_1 + a_2 + ... + a_k)^*$ \\
    \bottomrule
  \end{tabular}
\end{table}

\begin{table}[t]
  \vspace*{-0.2cm}
  \setlength{\tabcolsep}{5.5pt} 
  \caption{Number of result pairs per query.}
  \vspace*{-0.3cm}
  \label{tab:result}
  \small
  \begin{tabular}{cccccccc}
    \toprule
    \textbf{SF} & \textbf{Q1} & \textbf{Q2} & \textbf{Q3} & \textbf{Q4} & \textbf{Q5} & \textbf{Q6} & \textbf{Q7}\\
    \midrule
     1 & \multicolumn{1}{r}{105.6\,M} & \multicolumn{1}{r}{1.2\,B} & \multicolumn{1}{r}{1.0\,B} & \multicolumn{1}{r}{0.5\,M} & \multicolumn{1}{r}{4.2\,B} & \multicolumn{1}{r}{137.1\,M} & \multicolumn{1}{r}{39.4\,B} \\
     10 & \multicolumn{1}{r}{4.9\,B} & \multicolumn{1}{r}{53.2\,B} & \multicolumn{1}{r}{48.3\,B} & \multicolumn{1}{r}{3.6\,M} & \multicolumn{1}{r}{72.2\,B} & \multicolumn{1}{r}{930.7\,M} & \multicolumn{1}{r}{2.4\,T} \\
    \bottomrule
  \end{tabular}
  \vspace*{-0.2cm}
\end{table}

\begin{figure}[htbp]
\vspace*{-0.2cm}
    \centerline{\includegraphics[width=3.2in]{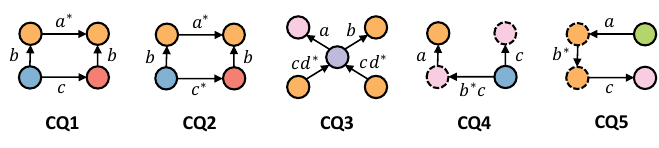}}
    \vspace*{-0.35cm}
    \caption{CRPQ queries used in the experiments.}
    \label{fig:CRPQ_graph}
    \vspace*{-0.2cm}
\end{figure}

\vspace{0.1cm}
\textbf{CRPQ datasets and queries:}
We evaluate CRPQ performance using the five queries\,(CQ1-CQ5) shown in Figure~\ref{fig:CRPQ_graph}.
The data graph is the same LDBC SNB dataset used for the RPQ experiments, and we consider two scale factors: SF=1 and SF=10.
The queries are derived from the LDBC microbenchmark \textit{Large-scale Complex Subgraph Query Benchmark}~\cite{mhedhbi2021lsqb}, which focuses on conjunctive queries.
We extend some edges by incorporating transitive closure, transforming the original conjunctive queries into CRPQs.
Vertex labels are distinguished by colors, and edges are annotated with the corresponding regular expressions.
CQ4 and CQ5 include a pair of vertices represented by dashed lines, which indicate a filter condition that both vertices must map to\,distinct\,vertices\,in\,the\,data\,graph.

\textbf{Systems compared:}
We compare cuRPQ with state-of-the-art methods.
First, we compare with DuckDB~\cite{raasveldt2019duckdb} and Umbra~\cite{neumann2020umbra}, which are CPU-based relational DBMSs following an algebra-based approach.
We also compare with Ring-RPQ~\cite{arroyuelo2024optimizing}, a CPU-based system that follows an automata-based approach.
Although Ring-RPQ supports \textit{reverse} execution plan\,(e.g., $A_1$ in Figure~\ref{fig:WavePlan}), it does not support parallel execution.
Therefore, we extend its implementation so that each CPU core can execute independently from different starting vertices.
Table~\ref{tab:system} summarizes the systems compared in our experiments.

Meanwhile, several well-known graph processing systems that support RPQ processing are excluded from this comparison.
For example, Virtuoso~\cite{erling2012virtuoso} and Blazegraph~\cite{thompson2016bigdata}\,(CPU-based RDF store DBMSs), as well as TigerGraph~\cite{deutsch2019tigergraph}\,(a CPU-based graph DBMS), are excluded because all-pairs RPQ evaluation leads to exponential path growth and prohibitively long execution times.
In addition, Kùzu~\cite{feng2023kuzu}\,(a CPU-based graph DBMS), HeavyDB~\cite{heavydb2025}\,(a GPU-based relational DBMS), and RAPIDS~\cite{RAPIDS2025}\,(a GPU-based relational library) do not support transitive closure~\cite{mulder2025optimizing}.
We provide a more detailed comparison of these systems in Section~\ref{sec:experiment_RPQ}, including their performance on small-batch RPQ.

\begin{table}[t]
    \vspace*{-0.1cm}
  \caption{Summary of characteristics of compared systems.}
  \vspace*{-0.3cm}
  \label{tab:system}
  \small
  \begin{tabular}{cccccc}
    \toprule
    \textbf{Approach} & \textbf{\centering Systems} & \textbf{\parbox{0.8cm}{\centering CPU/ GPU}} & \textbf{\parbox{1.1cm}{\centering Support CRPQ}} & \textbf{\parbox{1.6cm}{\centering In-memory/ Disk-based}} \\
    \midrule
    \multirow{2}{*}{\centering Algebra-based}
    & DuckDB   & CPU & O & Disk-based \\
    & Umbra    & CPU & O & Disk-based \\
    \midrule
    \multirow{2}{*}{Automata-based}
    & Ring-RPQ & CPU & X & In-memory \\
    & cuRPQ  & GPU & O & Disk-based \\
    \bottomrule
  \end{tabular}
  \vspace*{-0.2cm}
\end{table}

\textbf{HW/SW environment:}
We conduct all experiments on a dual-socket server equipped with two AMD EPYC 7302 processors\,(16 cores each), featuring a hierarchical cache architecture with a total of 1\,MiB L1d, 16\,MiB L2, and 256\,MiB L3 cache.
The system is configured with 1\,TB of DDR4 memory operating at 2933\,MT/s\,(DDR4-3200 DIMMs), four 6.4\,TB PCIe SSDs\,(RAID~0), and one NVIDIA A100 GPU with 80\,GB of memory.
For CPU-based methods, all CPU cores and memory are utilized without restrictions, and their default settings are applied.
For cuRPQ, unless otherwise specified, we allocate 24\,GB as GPU buffer memory, with 2\,GB for the input buffer, 20\,GB for the segment buffer, and 2\,GB for the UR buffer.
The operating system is Ubuntu 20.04.02, with g++ 7.5.0 as the compiler,\,and CUDA Toolkit 11.6 for kernel compilation.
The versions of the systems compared in the experiments are DuckDB 1.2.2 and Umbra 30b000783.
To ensure fair comparisons, we evaluate the performance of all methods with memory and OS caches cleared before each execution\,(i.e., cold start).
The only exception is Ring-RPQ, which operates in an in-memory fashion after loading the entire graph into main memory.
Accordingly, graph loading time from SSDs to main memory is excluded, and only query execution time is measured.

\vspace{-0.1cm}
\subsection{Comparison of RPQ Performance}
\label{sec:experiment_RPQ}

\begin{figure*}
\vspace*{-0.2cm}
    \centerline{\includegraphics[width=7.2in]{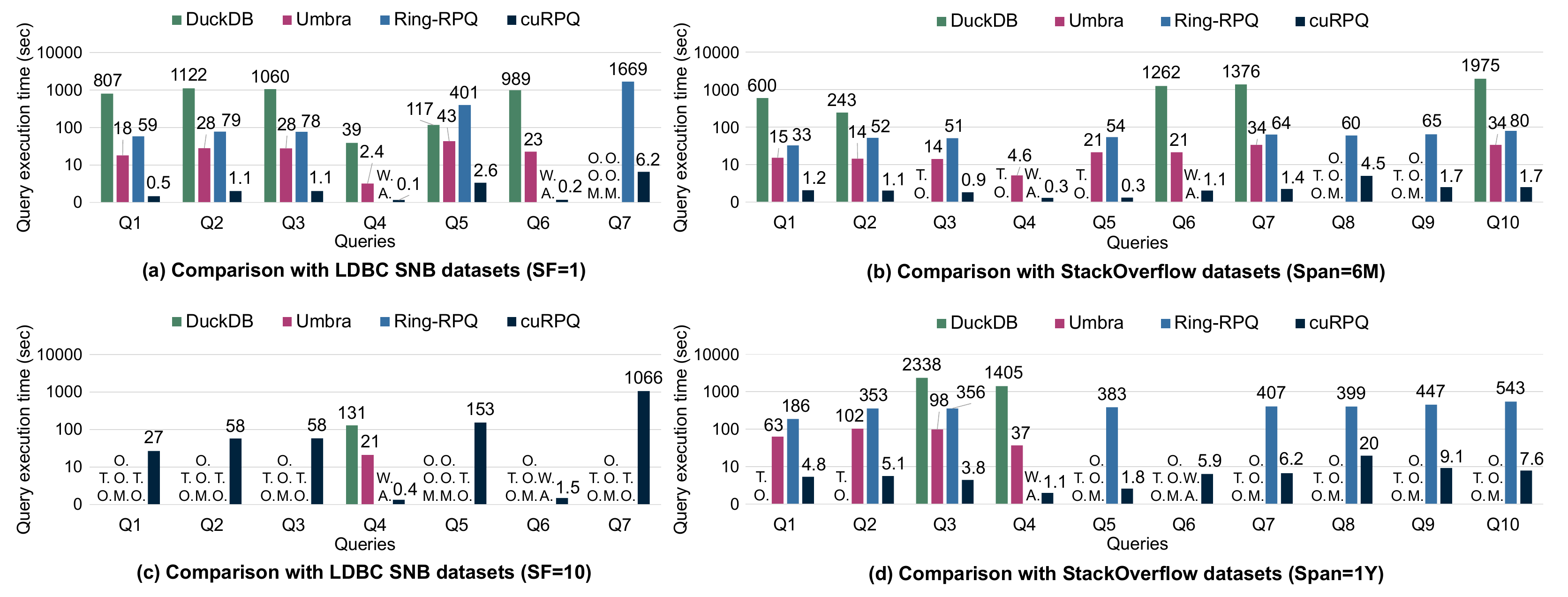}}
    \vspace*{-0.4cm}
    \caption{Comparison of RPQ execution times\,(T.O.: elapsed time exceeds 1 hour, O.O.M.: out of memory, W.A.: wrong answer).}
    \label{fig:ex1}
    \vspace*{-0.2cm}
\end{figure*}

Figure~\ref{fig:ex1} shows the query execution times of DuckDB, Umbra, Ring-RPQ, and cuRPQ on the LDBC SNB and StackOverflow datasets.

\vspace{0.1cm}
\textbf{Comparison with algebra-based systems:}
For the LDBC SNB dataset, both DuckDB and Umbra fail to complete Q7 at SF=1, and they fail to process all queries except for the non-recursive query Q4 at SF=10.
DuckDB encounters O.O.M. errors for Q5 and Q7 at SF=1 due to the large number of vertices to be explored, while the remaining queries terminate with T.O.
Umbra shows overall better performance than DuckDB, but execution still aborts due to O.O.M.
This indicates a limitation of algebra-based approaches, where the materialization of intermediate results by the $\alpha$-operator can cause O.O.M. or excessive computational overhead leading to T.O.
A similar limitation is also observed on the StackOverflow dataset.
Interestingly, DuckDB fails with timeouts for Q3 and Q4 when Span=6M, but it completes without timeouts on the larger dataset\,(Span=1Y) due to differences in the execution plan.
In contrast, cuRPQ successfully processes all queries. 
cuRPQ outperforms DuckDB by up to 4,945X\,(Q6 at SF=1), and Umbra by up to 115X\,(Q6\,at\,SF=1).

\vspace{0.1cm}
\textbf{Comparison with automata-based systems:}
Ring-RPQ fails to return correct results for specific queries such as Q4, Q6, and Q7 due to implementation issues.
For evaluation, we allow errors within a 3\% margin, where the \textit{error rate} represents the proportion of result pairs that a system fails to find with respect to the ground truth.
If the error rate exceeds this threshold, we denote the result as W.A.
Ring-RPQ generally performs slower than Umbra but faster than DuckDB, except for Q5 at SF=1.
Automata-based approaches can propagate results directly through state transitions in the automaton, thus enabling pipelined execution.
Thanks to this property, Ring-RPQ can process queries without O.O.M. for Span=1Y, in contrast to algebra-based systems.
However, at SF=10, the number of paths grows exponentially, leading all queries to terminate with T.O. or W.A.
In comparison, cuRPQ outperforms Ring-RPQ by up to 269X\,(Q7 at SF=1).

\begin{figure}[t]
\vspace*{-0.2cm}
    \centerline{\includegraphics[width=3.4in]{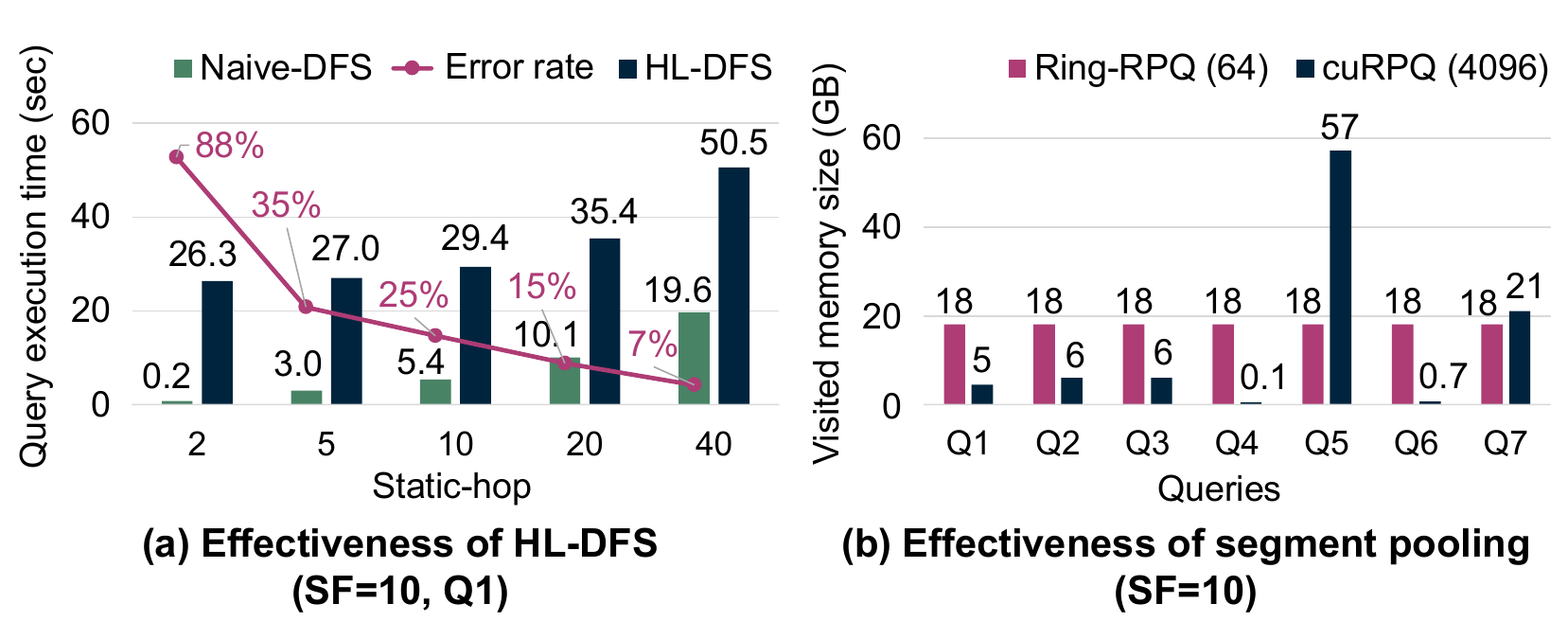}}
    \vspace*{-0.4cm}
    \caption{Impact of HL-DFS and segment pooling.}
    \label{fig:ex2}
    \vspace*{-0.3cm}
\end{figure}

\vspace{0.1cm}
\textbf{Performance overhead of arbitrary path lengths in RPQ:}
Table~\ref{tab:max_hop} compares cuRPQ’s Hop-limited Level-wise DFS\,(denoted as \textit{HL-DFS}) with a method that handles arbitrary paths using DFS with a sufficiently large static-hop limit\,(denoted as \textit{Naive-DFS}).
When we vary the static-hop size, we measure the maximum hop depth reached during exploration.
With a static-hop size of 2, HL-DFS explores up to 6 hops and finds all results.
In contrast, even when Naive-DFS increases the static-hop size up to 40 --- the maximum allowed in GPU and shared memory --- it still yields an error rate of 7.1\%.
This is because DFS explores paths deeply, often leading to exploration of unnecessarily long paths.
HL-DFS performs DFS within each TG, so expansion-TGs are still needed for long paths, though larger static-hop sizes reduce the number of required expansions\,(e.g., three expansions at static-hop=2 and two at 40).
Figure~\ref{fig:ex2}(a) presents the query execution times of both methods.
Both Naive-DFS and HL-DFS show increased execution time as the static-hop size grows due to their DFS nature, but HL-DFS can maintain efficiency with a reasonably small static-hop size.
In our experiments, we set the static-hop size of cuRPQ to 5.

\vspace{0.1cm}
\textbf{Memory overhead from visited set maintenance:}
Figure~\ref{fig:ex2}(b) shows the maximum memory consumption of the visited set for two automata-based systems, Ring-RPQ and cuRPQ.
Ring-RPQ processes up to 64 starting vertices simultaneously by leveraging 64 CPU virtual cores, whereas cuRPQ sets the batch size to 4,096 to exploit thousands of GPU cores.
Ring-RPQ implements the visited set using a wavelet tree data structure that supports simultaneous state transitions, so its memory usage is theoretically somewhat larger but remains constant.
In contrast, cuRPQ manages only the segments relevant to the query.
Even when processing 4,096 starting vertices simultaneously --- 64 times more than Ring-RPQ --- it achieves significantly lower memory usage\,(180X lower for Q4).
For Q5 and Q7, the memory demand approaches the segment buffer limit\,(20\,GB), but buffer consumption is effectively controlled within 20\,GB through sub-TG partitioning with bridge segments.

\begin{figure}[b]
\vspace*{-0.2cm}
    \centerline{\includegraphics[width=3.4in]{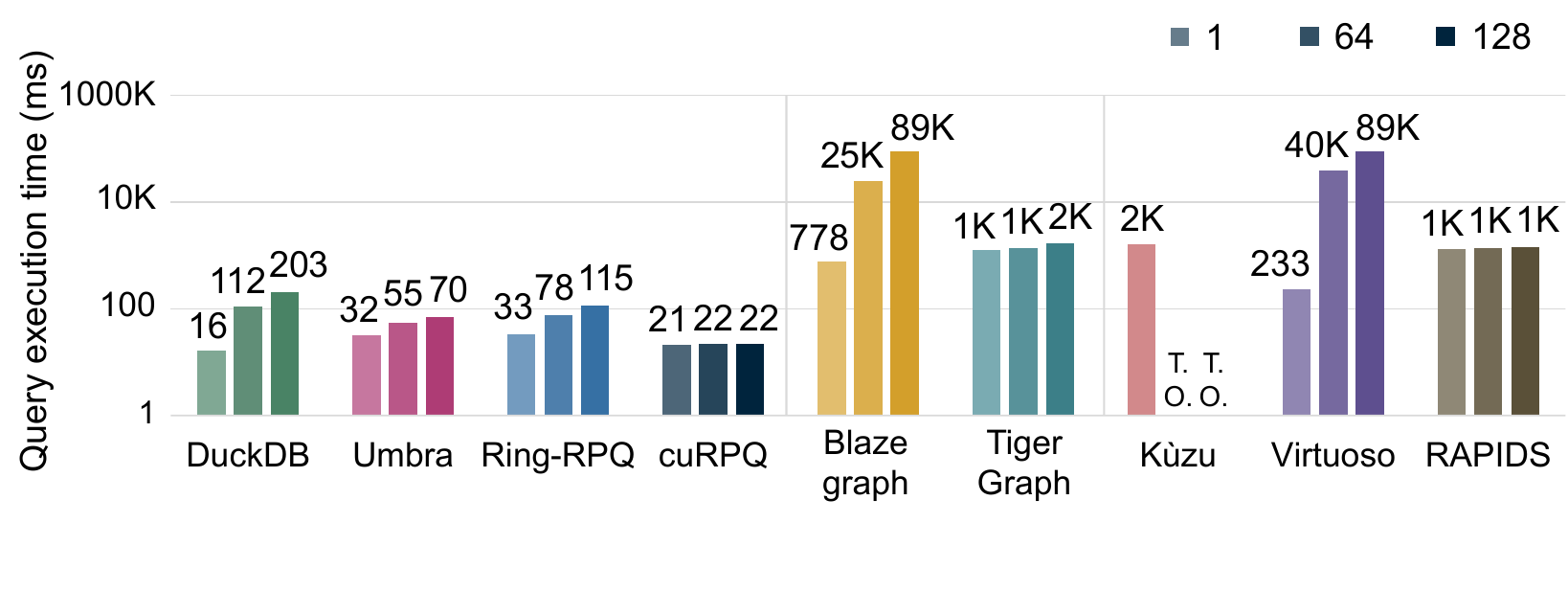}}
    \vspace*{-0.8cm}
    \caption{Comparison of query execution times on small-batch RPQ\,(LDBC SNB, SF=0.1, Q1).}
    \label{fig:ex3}
    \vspace*{-0.1cm}
\end{figure}

\vspace{0.1cm}
\textbf{Comparison with small-batch RPQ:}
Figure~\ref{fig:ex3} shows the query execution times of various systems that support RPQ when the number of starting vertices is 1, 64, and 128.
To provide a more precise comparison, we conducted the evaluation on a smaller dataset with SF=0.1\,(0.4M vertices and 2.1M edges).
The timeout threshold is set to 5 minutes.
The system versions used for comparison are Blazegraph 2.1.6, TigerGraph Enterprise Edition 3.10.4, Kùzu 0.4.2, Virtuoso 7.2.15, and RAPIDS 24.12.
For Kùzu and RAPIDS, which do not support the transitive closure, we limited the maximum hop length to 4, which may cause some paths\,(e.g., those reachable only through five hops) to be missed.
Virtuoso supports transitive closure but lacks duplicate detection, leading to excessive redundant exploration and ultimately T.O.
Thus, we limited Virtuoso to a maximum hop length of 4.
HeavyDB does not support the $\alpha$-operator or UNION operator, and is excluded from the comparison.

\begin{table}[t]
  \caption{Maximum hop counts with varying static-hop size.}
  \vspace*{-0.25cm}
  \label{tab:max_hop}
  \small
  \begin{tabular}{
    >{\centering\arraybackslash}m{1.3cm}
    *{5}{>{\centering\arraybackslash}m{0.6cm}}
    >{\centering\arraybackslash}m{1.7cm}
    }
    \toprule
    
    \multirow{2}{*}{\textbf{Search}} & \multicolumn{5}{c}{\textbf{Maximum hops per static-hop size}} & \multirow{2}{*}{\parbox{1.7cm}{\centering \textbf{Error rate  range }}}  \\
    \cline{2-6}
     & \vspace*{-0.05cm}\centering\makecell{\textbf{2}} & \textbf{5} & \textbf{10} & \textbf{20} & \textbf{40} & \\
    \midrule
     Naive-DFS & $\leq 2$ & $\leq 5$ & $\leq 10$ & $\leq 20$ & $\leq 40$ & $7.1\,\%\sim87.9\,\%$ \\
     HL-DFS & $\leq 6$ & $\leq 15$ & $\leq 20$ & $\leq 40$ & $\leq 80$ & $0\,\%$ \\
    \bottomrule
  \end{tabular}
  \vspace*{-0.2cm}
\end{table}

Systems other than the four main baselines considered in this paper show performance between 11X slower\,(Virtuoso) up to 4,045X slower\,(Blazegraph and Virtuoso) compared to cuRPQ.
Blazegraph is known to internally optimize transitive closure operations, resulting in relatively faster performance for single starting vertices, but its performance degrades sharply as the number of starting vertices increases.
RAPIDS efficiently utilizes the GPU, showing little to no performance degradation as the number of starting vertices increases.
However, it can only process datasets that fit into GPU memory, and O.O.M. occurs for all-pairs RPQ.
Although cuRPQ shows lower performance than DuckDB for single-source RPQ because only one thread block is used, it achieves substantial performance improvements as the number of starting vertices increases and the workload approaches all-pairs RPQ by exploiting massive GPU parallelism\,(9.2X at 128 vertices).

\vspace{-0.2cm}
\subsection{Comparison of CRPQ Performance}
\label{sec:experiment_CRPQ}

\begin{figure}[t]
\vspace*{-0.1cm}
    \centerline{\includegraphics[width=3.4in]{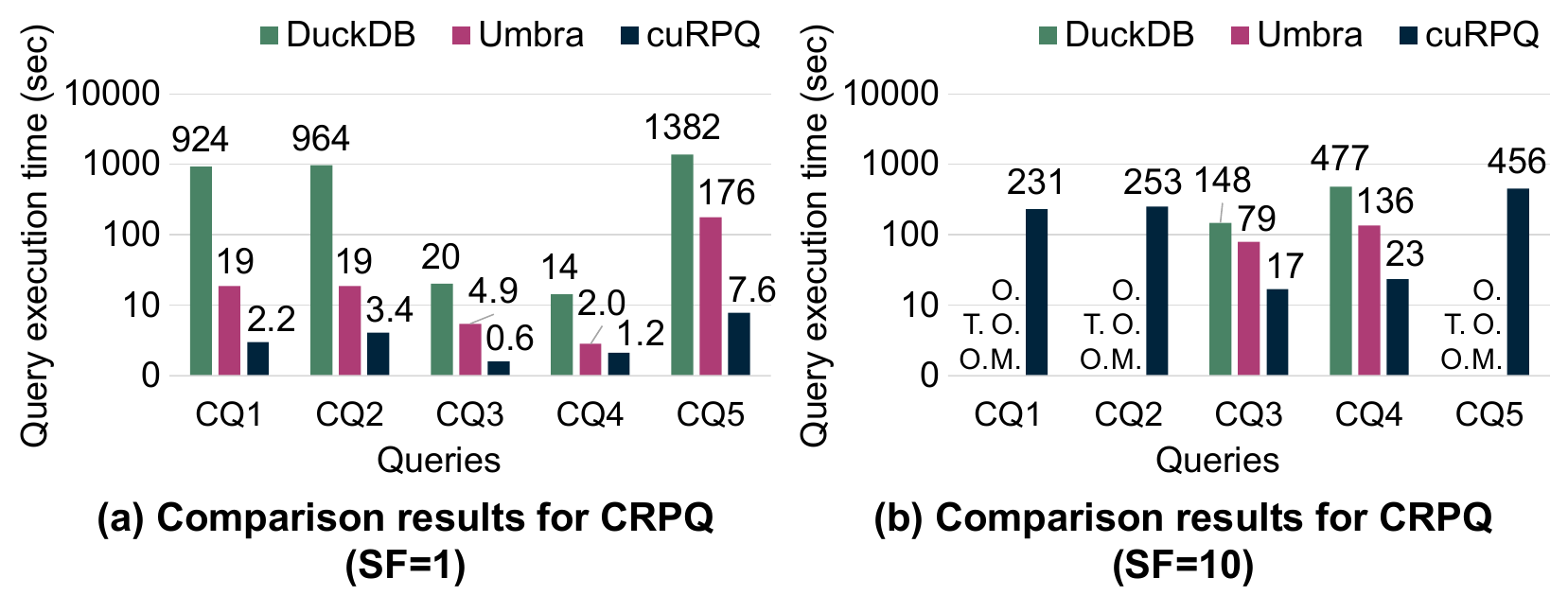}}
    \vspace*{-0.3cm}
    \caption{Comparison of CRPQ execution times.}
    \label{fig:ex4}
\end{figure}

Figure~\ref{fig:ex4} shows the query execution times of CRPQ queries on DuckDB, Umbra, and cuRPQ.
CQ1 incorporates Q1 as an RPQ atom.
Since algebra-based systems DuckDB and Umbra inevitably materialize intermediate results even for standalone RPQs, the additional cost of executing CQ1 compared to Q1 is negligible\,(e.g., 18\,sec and 19\,sec with Umbra).
In contrast, cuRPQ incurs a modest overhead due to BIM\,(0.5\,sec and 2.2\,sec), yet it still outperforms DuckDB by up to 420X\,(CQ1 at SF=1) and Umbra by up to 23.2X\,(CQ5 at SF=1).
The same trend is observed at SF=10.
DuckDB and Umbra fail to execute CQ1, CQ2, and CQ5 due to T.O. and O.O.M., respectively, when performing transitive closure.
cuRPQ successfully processes all queries and outperforms DuckDB by 20.7X and Umbra by 5.9X for CQ4.

\begin{figure}[t]
\vspace*{-0.2cm}
    \centerline{\includegraphics[width=3.4in]{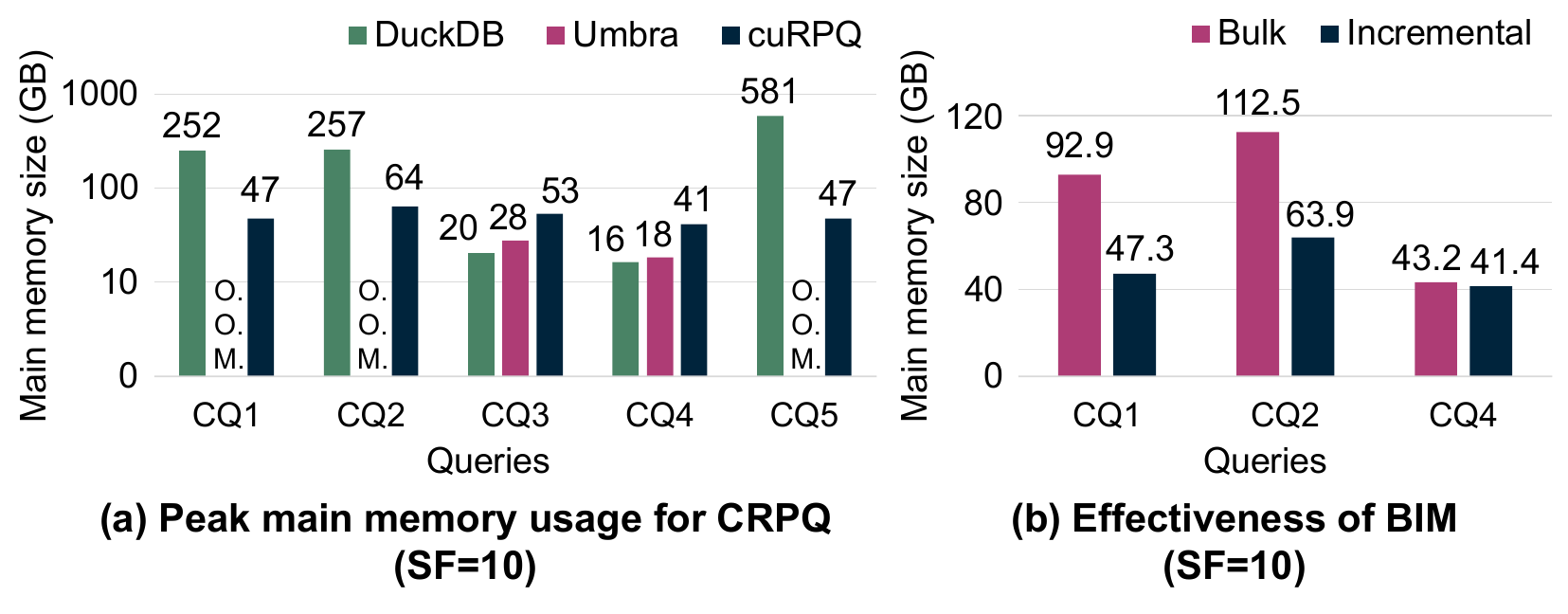}}
    \vspace*{-0.3cm}
    \caption{Intermediate data sizes in CRPQ.}
    \label{fig:ex5}
    \vspace*{-0.3cm}
\end{figure}

\vspace{0.1cm}
\textbf{RPQ result explosion for CRPQ:}
Figure~\ref{fig:ex5}(a) shows the peak main memory consumption during CRPQ execution.
At SF=10, DuckDB encounters T.O. for CQ1, CQ2, and CQ5, but we report the peak memory usage observed within the 2-hour time limit.
Umbra exhausts the entire 1\,TB of memory on these three queries.
In contrast, cuRPQ consumes up to 12.4X less memory than DuckDB for CQ5 and at least 21.3X less memory than Umbra for CQ1 and CQ5.
For CQ3 and CQ4, cuRPQ uses slightly more memory, as it must maintain block-level temporary slice buffers that have not yet been materialized.
However, cuRPQ immediately releases completed temporary slice buffers through BIM and keeps only the compressed graph format in main memory.
Thus, cuRPQ effectively controls main memory usage within 64\,GB even for queries such as CQ1, CQ2, and CQ5, where RPQ result explosion is particularly severe.

Figure~\ref{fig:ex5}(b) shows the peak memory consumption when results are materialized in bulk without BIM.
Without BIM, all RPQ results from the UR buffer must be retained in main memory in enumerated form, causing memory usage to scale with the output size of the RPQ atoms in CRPQ queries.
With BIM, memory consumption is reduced by up to 2.0X for CQ1.

\vspace*{-0.1cm}
 \subsection{GPU Performance and Parallelism Analysis}
\label{sec:analysis}

Table~\ref{tab:gpu_util_metrics} reports the \textbf{GPU utilization metrics} of cuRPQ.
The number of active warps per scheduler ranges from approximately 38-46\%,
while the number of active threads per warp varies between 67-75\%.
Given the inherently irregular control flow of graph traversal workloads, these values fall within a reasonable range~\cite{xie2025gpu}.
Meanwhile, DRAM throughput remains relatively low, whereas the L2 cache hit rate consistently stays around 90\%.
This indicates that RPQ evaluation in cuRPQ is largely compute-dominant rather than memory-bandwidth-bound, making it well suited for GPU acceleration.

Table~\ref{tab:parallelism_metrics} summarizes the characteristics of \textbf{query-level parallelism} in cuRPQ.
Depending on the query, the number of generated TGs exceeds one thousand, enabling effective coarse-grained parallel execution at the TG level.
The TG depth reflects the depth of the TG hierarchy formed by successive expansions during the hop-limited level-wise DFS.
As TG depth increases, the maximum reachable hop length is progressively extended according to the static-hop configuration; for example, Q8 at Span=1Y reaches up to $6 \times 5 = 30$ hops.
Across queries, the fan-out in base-TGs and the TG depth exhibit diverse patterns.
This diversity enables cuRPQ to accommodate both wide and deep traversal behaviors, effectively covering the arbitrary path lengths inherent in RPQ queries.

Table~\ref{tab:breakdown_bim} shows the \textbf{breakdown of BIM execution time}.
The overlap ratio measures the degree to which GPU-based exploration and CPU-side materialization are executed concurrently, normalized by the maximum possible overlap~\cite{denis2016mpi}.
Although the dominant cost varies across queries\,(e.g., materialization in CQ1 and exploration in CQ3 and CQ4), a substantial fraction of GPU and CPU work is overlapped.
These results demonstrate that BIM effectively reduces overall execution time by overlapping heterogeneous computation.

\begin{table}[t]
  \vspace*{-0.1cm}
  \caption{GPU utilization metrics\,(SF=10).}
  \vspace*{-0.3cm}
  \label{tab:gpu_util_metrics}
  \small
  \begin{tabular}{
    >{\centering\arraybackslash}m{0.8cm}
    >{\centering\arraybackslash}m{1.7cm}
    >{\centering\arraybackslash}m{1.8cm}
    >{\centering\arraybackslash}m{1.5cm}
    >{\centering\arraybackslash}m{0.9cm}}

    \toprule
    \textbf{Query} & \textbf{Active warps per scheduler} & \textbf{Active threads per warp} & \textbf{DRAM throughput} & \textbf{L2 hit rates} \\
    \midrule
    Q1 & 46.1~\% & 75.0~\% & 0.3~\% & 93.3~\% \\
    Q2 & 45.8~\% & 72.1~\% & 0.7~\% & 90.2~\% \\
    Q3 & 45.9~\% & 72.3~\% & 0.8~\% & 89.8~\% \\
    Q6 & 38.2~\% & 67.5~\% & 0.6~\% & 89.7~\% \\
    \bottomrule
  \end{tabular}
  \vspace*{-0.2cm}
\end{table}

\begin{table}[t]
  \caption{Query-level parallelism characteristics.}
  \vspace*{-0.35cm}
  \label{tab:parallelism_metrics}
  \small
  \begin{tabular}{
    >{\centering\arraybackslash}m{1.0cm}
    >{\centering\arraybackslash}m{0.8cm}
    >{\centering\arraybackslash}m{0.7cm}
    >{\centering\arraybackslash}m{1.1cm}
    >{\centering\arraybackslash}m{1.3cm}
    >{\centering\arraybackslash}m{1.5cm}}

    \toprule
    \textbf{Dataset} & \textbf{Query} & \textbf{\#\,TGs} & \textbf{Max.\ TG depth} & \textbf{Max.\ hops} & \textbf{Fan-out in base-TGs} \\
    \midrule
    \multirow{2}{*}{SF=10}
    & Q1 & 37 & 3 & 15 & 18 \\
    & Q5 & 1,423 & 2 & 10 & 1,388 \\
    \midrule
    \multirow{3}{*}{Span=1Y}
    & Q1 & 28 & 4 & 20 & 9 \\
    & Q5 & 16 & 4 & 20 & 5 \\
    & Q8 & 255 & 6 & 30 & 66 \\
    \bottomrule
  \end{tabular}
  \vspace*{-0.2cm}
\end{table}

\begin{table}[t]
  \caption{Breakdown of BIM execution time\,(SF=10).}
  \vspace*{-0.35cm}
  \label{tab:breakdown_bim}
  \small
  \begin{tabular}{
    >{\centering\arraybackslash}m{0.8cm}
    >{\centering\arraybackslash}m{1.2cm}
    >{\centering\arraybackslash}m{1.2cm}
    >{\centering\arraybackslash}m{1.9cm}
    >{\centering\arraybackslash}m{1.7cm}}

    \toprule
    \textbf{Query} & \textbf{GPU\,(sec)} & \textbf{CPU\,(sec)} & \textbf{BIM total\,(sec)} & \textbf{Overlap ratio} \\
    \midrule
    CQ1 & 46.9 & \underline{205.75} & 206.34 & 98.8~\% \\
    CQ3 & \underline{7.78}  & 2.75   & 8.75   & 64.8~\% \\
    CQ4 & \underline{18.07} & 3.65   & 19.18  & 69.6~\% \\
    \bottomrule
  \end{tabular}
  \vspace*{-0.2cm}
\end{table}

\vspace*{-0.1cm}
\subsection{Ablation Study on cuRPQ}
\label{sec:characteristics}

\begin{figure}[t]
\vspace*{-0.1cm}
    \centerline{\includegraphics[width=3.4in]{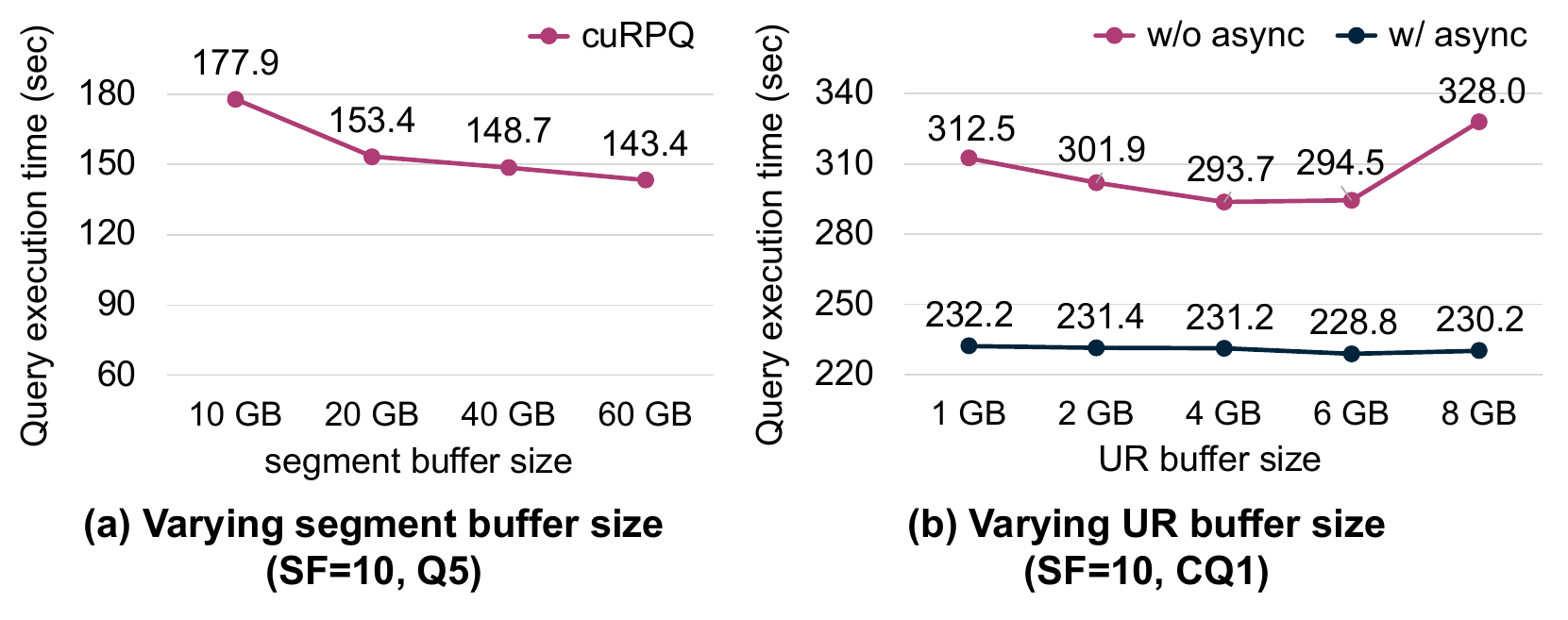}}
    \vspace*{-0.3cm}
    \caption{Impact of varying GPU buffer sizes.}
    \label{fig:ex6}
    \vspace*{-0.3cm}
\end{figure}

Figure~\ref{fig:ex6}(a) shows the query execution times with \textbf{varying segment buffer sizes}.
As shown earlier in Figure~\ref{fig:ex2}(b), Q5 at SF=10 requires up to 57.2\,GB of memory.
When the segment buffer size decreases, sub-TG partitioning occurs more frequently and the number of bridge segments to be managed increases, introducing additional overhead and gradually increasing execution time.
In particular, when the buffer size is reduced to as small as 10\,GB, segments cannot be released until all sub-TGs for a starting vertex have been processed, which can lead to insufficient segments in the pool.
In such cases, cuRPQ adopts the approach of temporarily reducing the batch size, which prevents full utilization of GPU parallelism and leads to performance degradation.

Figure~\ref{fig:ex6}(b) shows the query execution times with \textbf{varying UR buffer sizes}.
Since cuRPQ builds slices asynchronously through BIM, it shows only a negligible difference in execution time across UR buffer sizes.
If slice building were performed synchronously, the execution time would increase by up to 1.4X at 8\,GB compared to the asynchronous case.
In general, larger UR buffers improve performance by reducing the number of kernel suspensions and resumptions.
However, excessively large buffers transfer more results to the CPU at once, leading to the activation of more temporary slice buffers simultaneously.
This causes scattered writes and lowers cache efficiency, which in turn reduces performance.

\begin{figure}[t]
\vspace*{-0.2cm}
    \centerline{\includegraphics[width=3.4in]{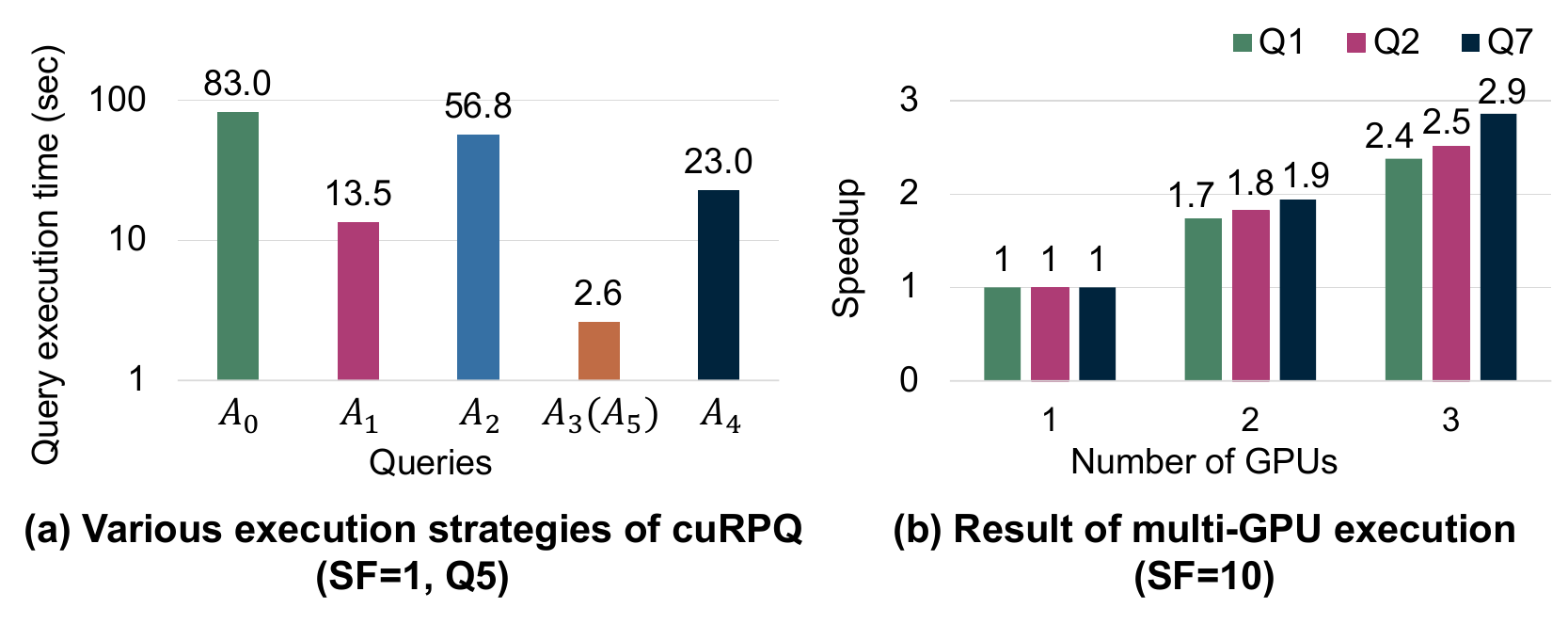}}
    \vspace*{-0.3cm}
    \caption{Scalability with execution strategies and GPUs.}
    \label{fig:ex7}
    \vspace*{-0.4cm}
\end{figure}

Figure~\ref{fig:ex7}(a) shows the execution time of \textbf{various execution strategies} applicable in cuRPQ.
For Q5\,($abc^*$), we evaluate plans $A_0$ through $A_4$, as shown in Figure~\ref{fig:AM_based}(a) and Figure~\ref{fig:WavePlan}.
Among these, $A_3$ and $A_4$ are start-in-the-middle plans. 
In cuRPQ, they are processed as modified plans by applying slice transpose, as explained in Figure~\ref{fig:CRPQ_and_execution_plans}(b).
$A_3$ is modified into $A_5$ and is denoted as ``$A_3(A_5)$''.
For Q5, the $A_3(A_5)$ plan is selected as the most efficient execution strategy, improving performance by up to 32X compared with $A_0$.

Figure~\ref{fig:ex7}(b) shows the \textbf{scalability of cuRPQ as the number of GPUs increases}.
cuRPQ executes batches of base-TGs concurrently across multiple GPUs, enabling scalable parallel processing.
Consequently, query performance scales almost linearly with the number of GPUs.
For three GPUs, workload imbalance causes longer idle times on some GPUs, slightly reducing efficiency for Q1 and Q2.

\vspace*{-0.1cm}
\section{Related Work}
\label{sec:related_work}

\begin{itemize}[labelindent=0em,labelsep=0.3em,leftmargin=*,itemsep=0em]

\item \textbf{Optimization techniques for RPQ evaluation:}
In automata-based approaches, RPQ evaluation has been optimized by decomposing regular expressions or selecting cost-effective sub-RPQs~\cite{koschmieder2012regular, nguyen2017efficient, arroyuelo2024optimizing}.
For multi-query optimization, techniques have been proposed to reduce computational cost by sharing common parts across multiple RPQs~\cite{na2022regular, abul2017multiple, abul2016swarmguide, pang2024materialized}.
Since cuRPQ provides a flexible plan representation, it can naturally incorporate these optimizations to achieve further performance\,improvements.

\item \textbf{Optimization techniques for transitive closure:}
Several approaches have accelerated transitive closure computation based on Warshall’s algorithm~\cite{warshall1962theorem} using Boolean algebra~\cite{katz2008all, arroyuelo2025evaluating, penn2006efficient}, and GPU algorithms have been explored to address bottlenecks from irregular memory access and thread synchronization~\cite{green2021anti, shovon2023towards}.
cuRPQ can flexibly integrate these techniques with the loop-cache plan to further improve transitive closure performance.

\item \textbf{Persistent RPQ in streaming graph processing:}
Persistent RPQs have been studied in streaming graphs, which monitor pre-registered queries and update results incrementally~\cite{gou2024lm, pacaci2020regular, pacaci2022evaluating}, commonly using datasets such as LDBC SNB and StackOverflow.
For LDBC SNB, evaluations adopt fine-grained windows of 3 days with sliding intervals of 6 hours~\cite{gou2024lm}.
Although cuRPQ does not directly support persistent RPQs, it efficiently processes large-scale snapshots\,(e.g., entire graphs or yearly intervals) and demonstrates strong performance even on challenging datasets.

\end{itemize}

\vspace*{-0.1cm}
\section{Conclusion and Discussion}
\label{sec:conclusion}

In this paper, we proposed cuRPQ, a GPU-optimized framework that preserves the automata-based exploration paradigm while efficiently supporting CRPQs, which are traditionally handled inefficiently by algebra-based methods.
To avoid the substantial data migration and synchronization overhead incurred by transparent GPU memory spilling during materialization, cuRPQ explicitly decouples GPU-based exploration from CPU-side materialization, enabling stable and high-throughput execution.
Extensive experiments demonstrated that cuRPQ successfully handled queries producing trillions of results, achieving up to 4,945X and 269X speedups over state-of-the-art algebra-based and automata-based methods, respectively.
Beyond its current design, cuRPQ can theoretically be extended to support RPQs with additional constraints such as length bounds and word-equality or prefix/suffix relations.
Length constraints can be naturally enforced by controlling traversal depth, while word-equality or prefix/suffix relations would require comparing label sequences across multiple paths during traversal.

\bibliographystyle{ACM-Reference-Format}
\bibliography{main}

\end{document}